\documentclass[%
preprint,
 amsmath,amssymb,
prb,
]{revtex4-2}

\usepackage{graphicx}
\usepackage{dcolumn}
\usepackage{bm}

\usepackage{xcolor}

\usepackage{mathtools}
\usepackage{braket}
\usepackage{hyperref}
\usepackage{bbold}
\usepackage{calrsfs}
\usepackage{dsfont}
\usepackage{xcolor}
\usepackage{subcaption}
\usepackage{float}

\newcommand{\iu}{{i\mkern1mu}}
\renewcommand{\vec}{\boldsymbol}

\begin{document}

\preprint{AIP/123-QED}

\title{ 
A protected spin-orbit induced absorption divergence in distorted Landau levels 
}
\author{Dominik Sidler}
  \email{dsidler@mpsd.mpg.de}
 \affiliation{Max Planck Institute for the Structure and Dynamics of Matter and Center for Free-Electron Laser Science, Luruper Chaussee 149, 22761 Hamburg, Germany}
\affiliation{The Hamburg Center for Ultrafast Imaging, Luruper Chaussee 149, 22761 Hamburg, Germany}

\author{Michael Ruggenthaler}
  \email{michael.ruggenthaler@mpsd.mpg.de}
  \affiliation{Max Planck Institute for the Structure and Dynamics of Matter and Center for Free-Electron Laser Science, Luruper Chaussee 149, 22761 Hamburg, Germany}
  \affiliation{The Hamburg Center for Ultrafast Imaging, Luruper Chaussee 149, 22761 Hamburg, Germany}

\author{Angel Rubio}
  \email{angel.rubio@mpsd.mpg.de}
  \affiliation{Max Planck Institute for the Structure and Dynamics of Matter and Center for Free-Electron Laser Science, Luruper Chaussee 149, 22761 Hamburg, Germany}
    \affiliation{The Hamburg Center for Ultrafast Imaging, Luruper Chaussee 149, 22761 Hamburg, Germany}
  \affiliation{Center for Computational Quantum Physics, Flatiron Institute, 162 5th Avenue, New York, NY 10010, USA}
  \affiliation{Nano-Bio Spectroscopy Group, University of the Basque Country (UPV/EHU), 20018 San Sebasti\'an, Spain}

\date{\today}

\begin{abstract}
The effect of spin-orbit (and Darwin) interaction on a 2D electron gas subject to a radial symmetric, inhomogeneous $1/r$-magnetic field is discussed analytically in a perturbative and non-perturbative manner. For this purpose, we investigate the radial Hall conductivity that emerges from an additional homogeneous electric field perturbation perpendicular to the 2D electron gas, which  solely interacts via spin-orbit coupling.   Numerical calculations of the absorptive spin-orbit spectra show   for an ideal InSb electron gas a behaviour that is dominated by the localized (atomic) part of the distorted Landau levels. In contrast, however, we also find analytically that a (non-local) divergent static response emerges for Fermi energies close to the ionization energy in the thermodynamic limit. The divergent linear response implies that the external electric field is entirely absorbed outside the 2D electron gas by induced radial spin-orbit currents, as it would be the case inside a perfect conductor. This spin-orbit induced polarization mechanism depends on the effective $g^*$-factor of the material for which it shows  a critical behaviour at $g^*_c=2$, where it abruptly switches direction.
The diverging absorption  relies on the presence of degenerate energies with allowed selection rules that are imposed by the radial symmetry of our inhomogeneous setup. We show analytically the presence of a discrete Rydberg-like band structure that obeys these symmetry properties. While in our case this structure turns out to be of  minor relevance, it is a promising property, which may facilitate the experimental realization in the future.
In a last step, we investigate the robustness of the spectra by solving analytically the Dirac equation expanded up to order $1/(mc)^2$. We find that the distorted Landau-levels, and thus the divergent spin-orbit polarization,  remain protected with respect to slow changes of the applied $1/r$-magnetic field.   
\end{abstract}

\maketitle

\section{Introduction}

The Hall effect has been a cornerstone of  solid state physics for almost 150 years since its first discovery.\cite{hall1879new}  The basic mechanism, that a charge currents get deflected in a perpendicular magnetic field, does not only hold on a macroscopic scale, but numerous seminal experiments have been designed, where delicate quantum features of the materials become apparent in all sorts of different Hall setups. For example, the  integer\cite{klitzing1980new} and fractional\cite{tsui1982two, Laughlingfractional} quantum Hall effect are potentially the most famous among those experiments. With growing experimental capabilities, more delicate experimental conditions have been realized over time, where even the quantization of the electro-magnetic fields can become decisive, as for example under ultra-strong coupling conditions in optical cavities.~\cite{ScalariScience, li2018, FaistCavityHall, paravacini2019,rubio2022new,Hagenmuller2010cyclotron, CiutiHopping, rokaj2019, RokajButterfly2021,de2021light} All of which experiments have in common that they fundamentally rely on the emergence of quantized Landau levels for a two dimensional electron gas, \cite{landau1930diamagnetismus, Landau} which provides a paradigmatic analytic model for materials in an (at least locally) homogeneous magnetic field. 

Only very recently, a simple analytic solution was provided for a non-interacting 2D electron gas in a truly inhomogeneous magnetic field, by assuming a rotationally symmetric and $1/r$-decaying magnetic field with respect to the distance $r$ from the origin.\cite{sidler2022class} It has been shown that this $B$-field ''impurity'' introduces strongly distorted Landau levels that can fundamentally alter physical observables, compared with Landau-level physics or the homogeneous electron gas.\cite{sidler2022class} When creating distorted Landau levels with a strong spin-dependency, it is now tempting to ask what spin-orbit effects could emerge in such a setting. In the subsequent manuscript we will uncover those basic mechanisms analytically and numerically for the spin-orbit Hall conductivity.      
Notice that our chosen setup is fundamentally different to the anomalous Hall effect in ferromagnetic materials\cite{RevModPhys.82.1539} or the related spin Hall effect,\cite{RevModPhys.87.1213} which also emerge from spin-orbit interaction and are of particular interest for the flourishing field of spintronics.\cite{RevModPhys.76.323} In those cases, three dominant mechanisms are known: The intrinsic scattering contributions\cite{karplus1954hall}, which emerge from the topological Fermi liquid property,\cite{haldane2004berry} and two extrinsic contributions (skew \cite{smit1955spontaneous,smit1958spontaneous} and side jump scattering\cite{berger1970side}) that are related to the scattering at a scalar potential well (e.g. charged impurity). All of these effects can jointly be described by Kubo's linear response theory,\cite{crepieux2001theory} which we will also rely on subsequently.  However, in  our case, the significant spin-orbit effects will not emerge from a scalar potential, instead they arise from the subtle interplay of a magnetic field-impurity (affecting the Zeeman interaction) with its corresponding vector-potential that couples to the momentum operator.

The  manuscript is structured as follows: First, we shortly recapitulate the recently introduced analytical solution for distorted Landau levels in an inhomogeneous magnetic field. Based on this, we determine  the radial Hall conductivity that emerges from spin-orbit interaction when applying a small electric field perturbation perpendicular to the 2D electron gas. In the next section, we focus on the static (DC) absorptive features of the radial Hall conductivity, which shows local as well as non-local (diverging) features depending on the Fermi energy. Furthermore, we discuss the influence of the effective $g^*$-factor on the results, which gives rise to different phases. Afterwards, we focus on the robustness of the previously derived results with respect to slow variations of the externally applied magnetic fields, by solving analytically the corresponding quasi-static eigenvalue problem. Finally, we summarize our results and assess them with respect to their potential experimental verification.  

\section{Radial Hall conductivity from spin-orbit interaction}

\begin{figure}
     \centering
     \begin{subfigure}[b]{0.3\textwidth}
         \centering
         \includegraphics[width=\textwidth]{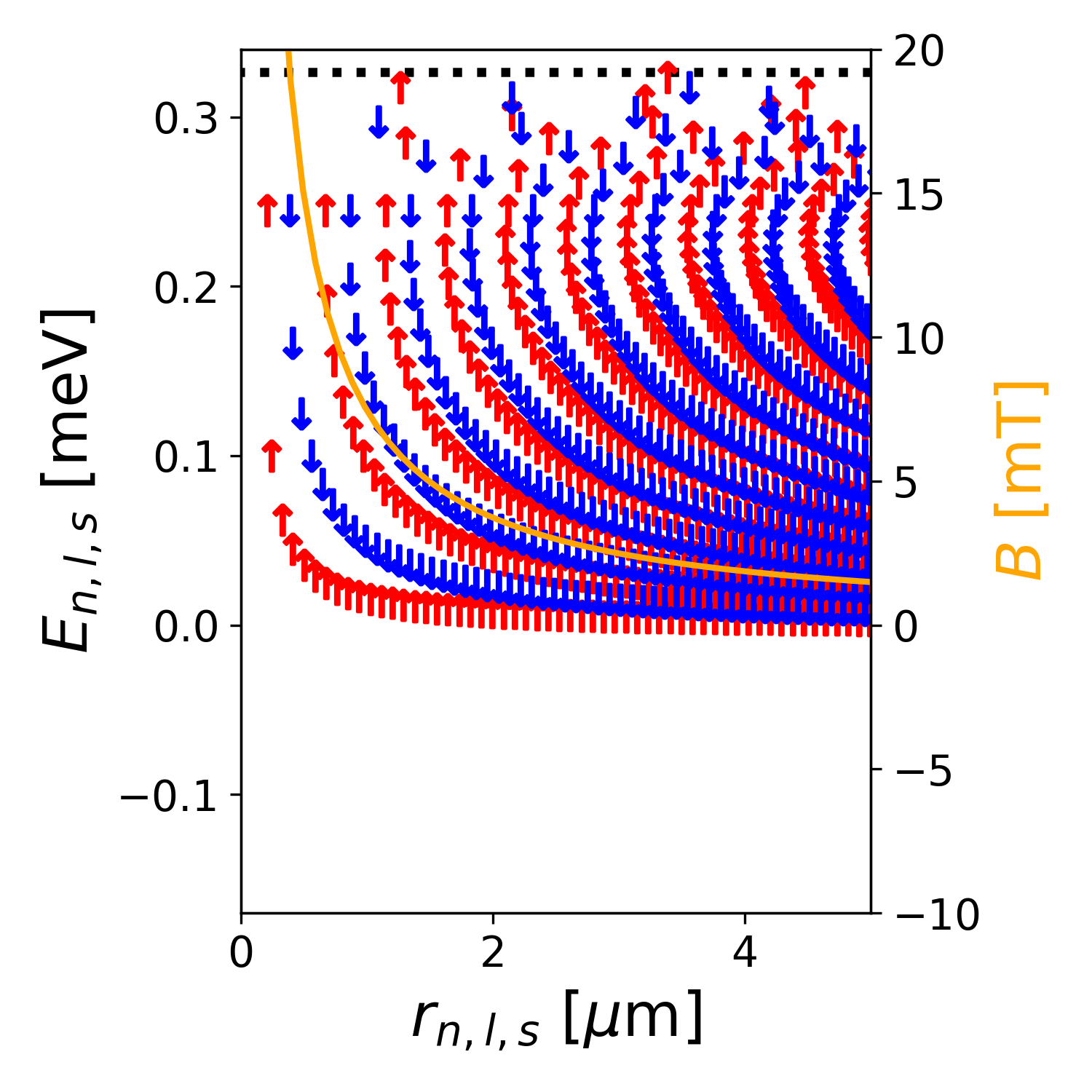}
         \caption{$g^*=1$}
         \label{fig:y equals x}
     \end{subfigure}
     \hfill
     \begin{subfigure}[b]{0.3\textwidth}
         \centering
         \includegraphics[width=\textwidth]{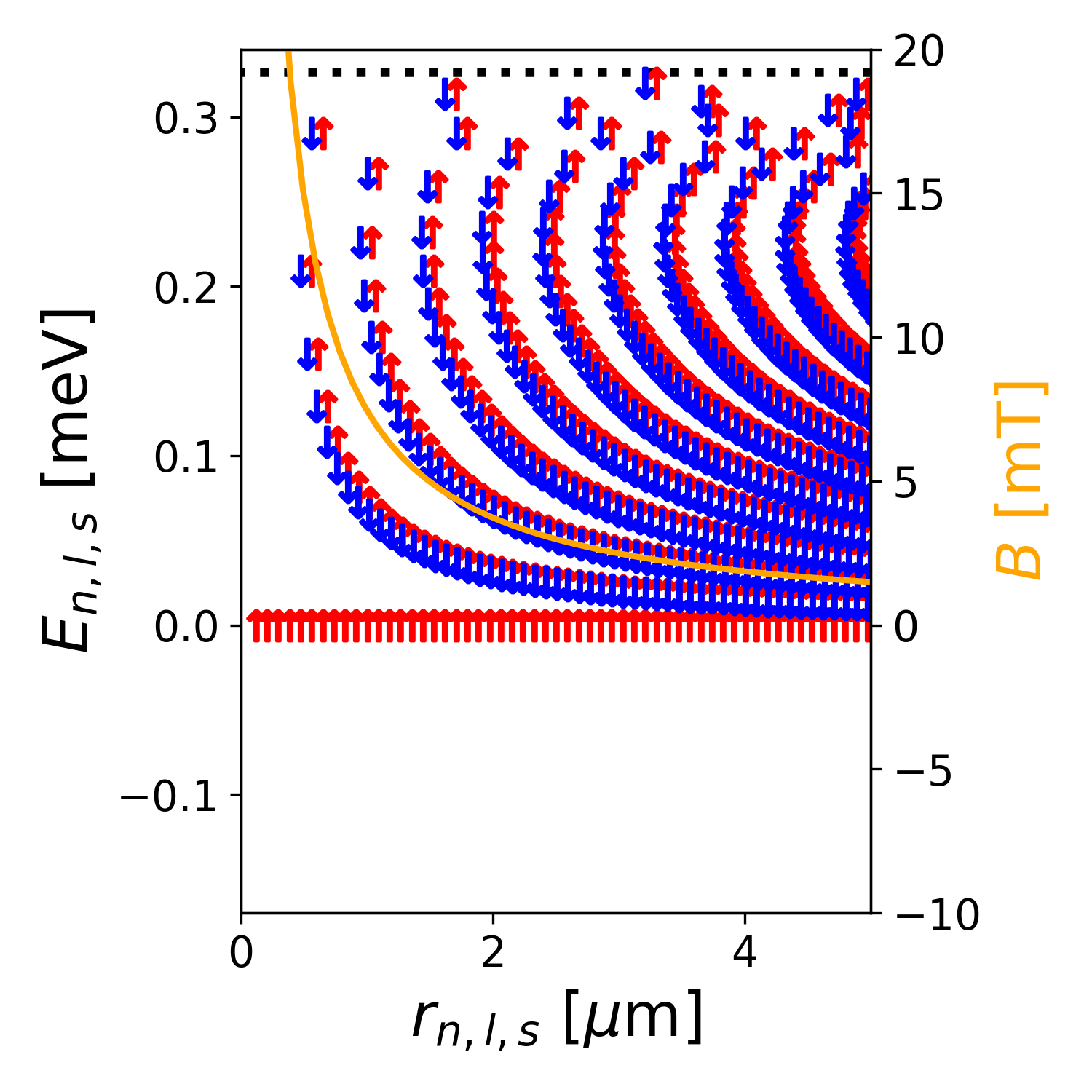}
         \caption{$g^*=2$}
         \label{fig:three sin x}
     \end{subfigure}
     \hfill
     \begin{subfigure}[b]{0.3\textwidth}
         \centering
         \includegraphics[width=\textwidth]{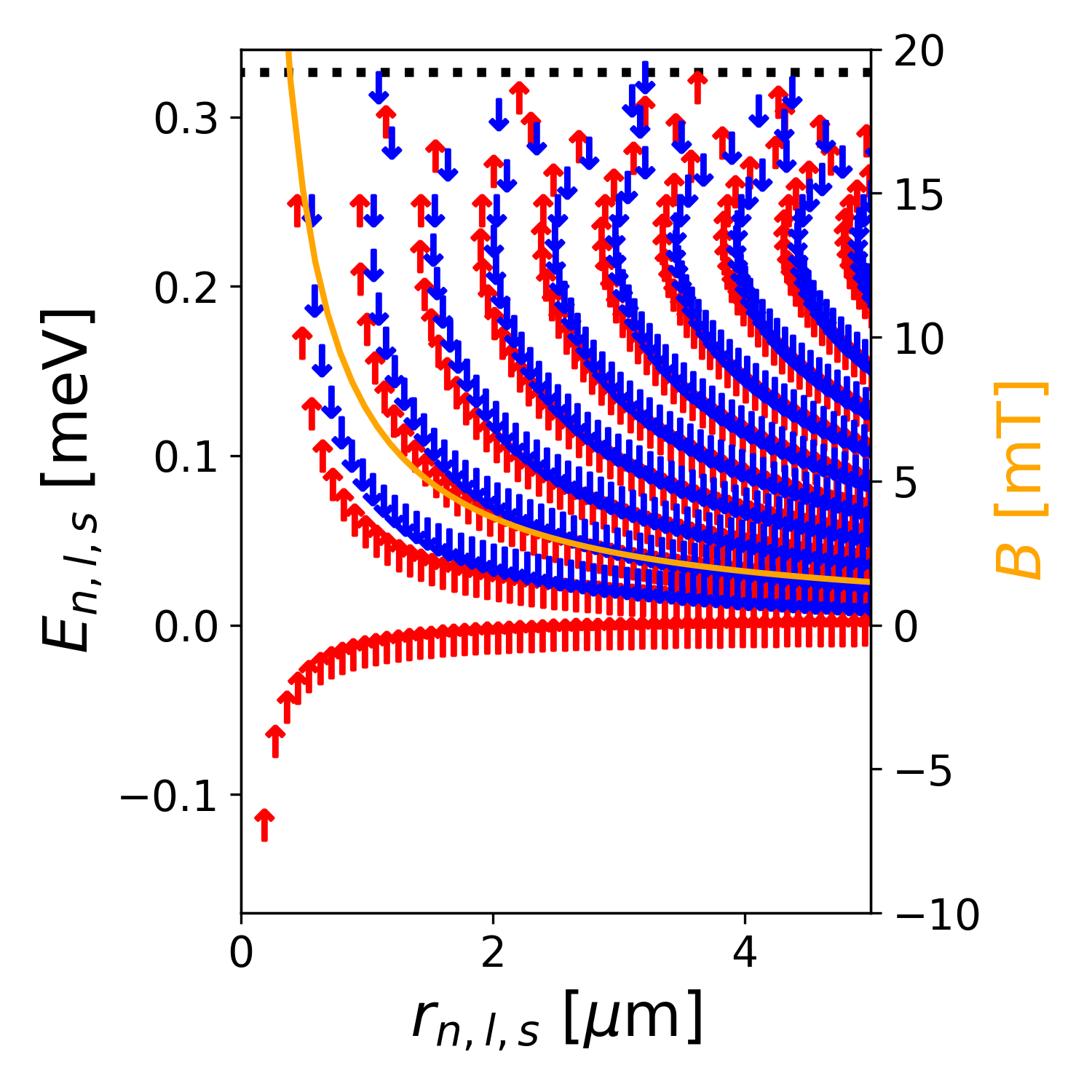}
         \caption{$g^*=3$}
         \label{fig:y equals x}
     \end{subfigure}\\
        \caption{Radially resolved bound state eigenvalue spectrum for an InSb 2D electron gas assuming different effective $g^*$ factors  and $A_\phi=-0.0006$ [a.u.].  A distorted Landau level structure becomes apparent caused by the applied $1/r$-decaying magnetic field (orange).\cite{sidler2022class} The up spin (red) and down spin (blue) dependencies are visualized as well as the ionization energy at $E_{A^2}=q^2 A_\phi^2/(2m^*)$ (dashed black line). An increase in the externally applied vector potential $A_\phi$ will result in an increase of the ionization threshold and simultaneously it will squeeze the states spatially more densely, i.e., it reduces the expected radial distance $r_{n,l,s}$ with respect to the origin.}
        \label{fig:distorted_level}
\end{figure}

The  starting point of our investigation of transversal spin-orbit Hall conductivity effects in a non-interacting 2D electron gas subject to a inhomogeneous magnetic field $\vec{B}(\boldsymbol{r})$, we use the Pauli Hamiltonian \cite{frohlich1993gauge,sidler2022class}
\begin{eqnarray}
\hat{H}_0&=&\sum_{j=1}^N\frac{\hat{\vec{\Pi}}_j^2}{2m^*}-\frac{g^*q\hbar}{4m^*} \hat{\vec{\sigma}}_j\cdot\vec{B}(\vec{r}_j).\label{eq:hamiltonian0}
\end{eqnarray}
The recently introduced bound state solution is given by,\cite{sidler2022class}
\begin{eqnarray}
E_{n,l,s}=\frac{q^2 A_\phi^2
}{2m^*}\bigg(1-\bigg[\frac{ 2l+g^* s}{2n+1}\bigg]^2\bigg),\ n\geq l,\ l+g^* s/2>0. \label{eq:Etotbound}
\end{eqnarray}
with Hydrogen-like eigenstates,
\begin{eqnarray}
    \Psi_{n,l,s}(t)=\frac{1}{\sqrt{N_{n,l,s}}}e^{il\phi}e^{-\frac{x(r)}{2}}x^l(r) L_{n-l}^{2l}(x(r))\chi(s),\label{eq:eigf}
\end{eqnarray}
which are written in terms of generalized Laguerre polynomials $L_w^\nu$. We have introduce an energy dependent radial scaling factor,
\begin{eqnarray}
x(r):=\frac{2 q A_\phi}{\hbar}\frac{2l+g^* s}{2n+1}r,\label{eq:scaling}
\end{eqnarray}
and the normalization constant,
\begin{eqnarray}
N_{n,l,s}&=&\int_0^{2\pi}\int_0^\infty \Psi_{n,l,s}^*\Psi_{n,l,s} rdr d\phi\nonumber\\
&=&2\pi \frac{(n+l)!}{(n-l)!}(2n+1)\bigg(\frac{2 q A_\phi}{\hbar}\frac{2l+g^* s}{2n+1}\bigg)^{-2}.
\end{eqnarray}
Notice that the inequality conditions for the allowed quantum numbers distinguishes the bound states from the unknown continuum solution, for which $l+g^* s/2\leq 0$ holds.\cite{sidler2022class} 
The canonical momentum operator is defined as $\hat{\vec{\Pi}}_j:=\hat{\vec{p}}_j-q \vec{A}(\boldsymbol{r}_j)$ and
the effective electron mass is indicated by $m^*$ with negative unit charge $q=-e$. The integer quantum numbers are labeled as $n, l$, where $l$ corresponds to the angular momentum quantum number. Furthermore, the electron spin-half quantum numbers are indicated by $s$. We denote the usual  position operator of particle $j$ as $\vec{r}_j$ and the corresponding momentum operator as $\hat{\vec{p}}_j$.  The anisotropic external vector potential is denoted by $\vec{A}(\boldsymbol{r})$ 
and an effective g-factor is introduced as  $g^*$, which in our definition explicitly excludes modifications due to the effective mass of the electrons, i.e. it only accounts for relativistic corrections or other higher order host material effects (e.g. influence of  Coulomb interaction or applied magnetic fields). Those mechanisms will not be discussed further. The Pauli vector for electron $j$ is labeled by $\vec{\sigma}_j$. 
In cylindrical coordinates the vector potential and corresponding magnetic field are given as,\cite{sidler2022class} 
\begin{eqnarray}
\vec{A}(\boldsymbol{r})&:=&A_\phi \vec{e}_{\phi}\\
 \vec{B}(\boldsymbol{r})&=&\vec{\nabla}\wedge \vec{A}=\frac{1}{r}\frac{\partial(r A_\phi)}{\partial r}\vec{e}_z=\frac{A_\phi}{r}\vec{e}_z.\label{eq:B}
\end{eqnarray}
such that $A_\phi$ is  constant, with $\vec{e}_{\phi}=\frac{1}{r}(-y \vec{e}_x+x\vec{e}_y)$ indicating the unit vector along $\phi$-direction.
Notice, our inhomogeneous magnetic field $ \vec{B}(\boldsymbol{r})$ corresponds to a radial external  current density of the form $J_{\mathrm{ext}}=\frac{\vec{\nabla}\wedge \vec{B}}{\mu_0}=\frac{A_\phi}{\mu_0}\frac{1}{ r^2} \vec{e}_\phi$ if considered in free space, i.e., it does not correspond to a magnetic monopole. Eq. (\ref{eq:Etotbound}) gives rise to  a distorted Landau level structure (see Fig. \ref{fig:distorted_level} for different $g^*$-factors), which introduces a plethora of features such as infinite degeneracies, charge or current oscillations as well as Hall conductivity phase transitions for in-plane static electric field perturbations which have been previously been discussed in Ref. \citenum{sidler2022class}. 
Notice also that  likewise,  hydrogen related spectra emerge for two dimensional magnetic quantum dots.\cite{downing2016magnetic,downing2016massless}

In contrast to established properties within the context of Hall conductivity for in plane perturbations, we will investigate the impact of a time-dependent, homogeneous electric field $\bold{E}_z(t)=E_z e^{-\iu \omega t} \vec{e}_z$ perpendicular (!) to the 2D electron gas. For this purpose, we apply Kubo's linear response formalism to our setup. The leading order in-plane effects from $\bold{E}_z(t)$  are induced by the spin-orbit (and Darwin) interaction (according to the Foldy-Wouthuysen expansion of the Dirac equation up to order $1/(m^* c)^2$). 
The corresponding time-dependent perturbing Hamiltonian can be written as, \cite{frohlich1993gauge}
\begin{eqnarray}
\hat{H}_1(t)&=&-\sum_{j=1}^N\frac{q\hbar}{8m^{*2}  c^2}\Big[(\hat{\vec{p}}_j-q \vec{A}(\boldsymbol{r}_j))\cdot\big(\hat{\vec{\sigma}}_j\times\bold{E}_z(t)\big)+\big(\hat{\vec{\sigma}}_j\times\bold{E}_z(t)\big)\cdot(\hat{\vec{p}}_j-q \vec{A}(\boldsymbol{r}_j))\Big]\label{eq:hpert}\\
&=&-\sum_{j=1}^N\frac{q\hbar}{4m^{*2}  c^2}\Big[\big(\sigma_y(\cos(\phi)\vec{e}_r-\sin(\phi)\vec{e}_\phi)-\sigma_x(\sin(\phi)\vec{e}_r+\cos(\phi)\vec{e}_\phi)\big)\Big]_j\cdot\hat{\vec{\Pi}}_jE_z(t)\label{eq:hpert}\\
&=:&-\hat{d}_1 E_z(t) 
\end{eqnarray}
where the Darwin contribution $\propto \mathrm{div} \vec{E}$ vanishes due to assuming a homogeneous electric field.
From Eq. (\ref{eq:hpert}) we immediately notice that the perturbing Hamiltonian starts to mix the two different spin channels by $\sigma_x,\sigma_y$. This has not been the case for our $\hat{H}_0$, which solely depends on the identity Pauli matrix $\sigma_z$ instead. This 
has important implications, since it ensures that the only radially induced current $\langle \hat{j}^{r}(t)\rangle$ must be induced by the spin-orbit current operator $\hat{\bold{j}}^{\rm SO}_j:=-\hbar q/(4m^{*2}c^2)(\hat{\vec{\sigma}}_j\times \vec{E})$\cite{tancogne2022effect} within Kubo's linear response formalism. In contrast, paramagnetic, diamagnetic and magnetization currents cannot induce radial currents perturbatively, since they are all spin-preserving operators. Notice, this property, i.e., $  \hat{j}^{r}=\hat{\bold{j}}^{\rm SO}\cdot \vec{e}_r$, is universally true for any  radially symmetric magnetic fields with a superimposed homogeneous electric field perturbation, which may simplify experimental realizations.
Applying the Kubo linear response relation, the total radial current can be written as
\begin{eqnarray}
\langle \hat{j}^{r}(t)\rangle&=&\langle\hat{j}^{r}(t)\rangle_0-\iu\int_{0}^t dt^\prime \big\langle[\hat{j}^{r}(t),\hat{H}_1(t)]\big\rangle_0\\
&=&-\int_{-\infty}^\infty dt^\prime \hat{\sigma}^{ rz}(t-t^\prime) E_z(t^\prime),
%
\end{eqnarray}
where the first term becomes exactly zero.\cite{sidler2022class}  The $\langle\rangle_0$ indicates a ground state depending at temperature $T$. Additionally we introduced the radial Hall conductivity,
\begin{eqnarray}
\sigma^{rz}(t-t^\prime)=-\iu \theta(t-t^\prime)\big\langle[\hat{j}^{r}(t),\hat{d}_1(t)]\big\rangle_0,
\end{eqnarray}
with unit step function $\theta(t)$.
Taking the Laplace transform with respect to time transforms the convolution integral into,
\begin{eqnarray}
j^{ r}(\omega,E_F,T)
&=&-\sigma^{ rz}(\omega,E_F,T)E_z(\omega).
\end{eqnarray}
The frequency dependent (AC) transversal conductivity can be written  at temperature $T$,\cite{ashcroft2022solid,stefanucci2013nonequilibrium}
\begin{eqnarray}
\sigma^{rz}(\omega,E_F,T)
&=&-\iu \sum_{a,b}f(E_a)(1-f(E_b))\bigg[
\frac{\bra{a}\hat{j}^{r}\ket{b}\bra{b}\hat{d}_1 \ket{a}}{E_a-E_b+\hbar \omega+\iu \Gamma}+\frac{\bra{a}\hat{d}_1\ket{b}\bra{b}\hat{j}^{r} \ket{a}}{ E_a-E_b-\hbar\omega-\iu \Gamma}\bigg],\label{eq:socond_initial}
\end{eqnarray}
with Fermi distribution $f(E_a):=1/(e^{(E_a-E_F)/k_B T}+1)$ and $\Gamma>0$.  Notice that we implicitly restrict our subsequent analysis to bound state currents / conductivties, since the continuum states giving rise to free currents are not known. However, restricting the $\sum_{a,b}$ to bound states will not affect the validity of our calculations, as we will see later. The frequency dependent absorptive (imaginary) part of the spin-orbit conductivity can be found by the identity $1/(x+\iu \Gamma)=P(1/x)-\iu\pi\delta(x)$,\cite{stefanucci2013nonequilibrium} for which we find
\begin{eqnarray}
\Im(\sigma^{rz}(\omega,E_F,T))
&=&\pi\iu \sum_{a,b}f(E_a)(1-f(E_b))\cdot\nonumber\\
&&\bigg[\frac{\bra{a}\hat{j}^{r}\ket{b}\bra{b}\hat{d}_1 \ket{a}\Gamma}{(E_a-E_b+\hbar \omega)^2+\Gamma^2}-\frac{\bra{a}\hat{d}_1\ket{b}\bra{b}\hat{j}^{r} \ket{a}\Gamma}{ (E_b-E_a+\hbar\omega)^2+ \Gamma^2}\bigg]\label{eq:imac}
\end{eqnarray}
assuming a Lorentzian lineshape with a finite lifetime $\Gamma>0$ for the $\delta$-distribution. 

\begin{figure}
     \centering
     \begin{subfigure}[b]{0.3\textwidth}
         \centering
         \includegraphics[width=\textwidth]{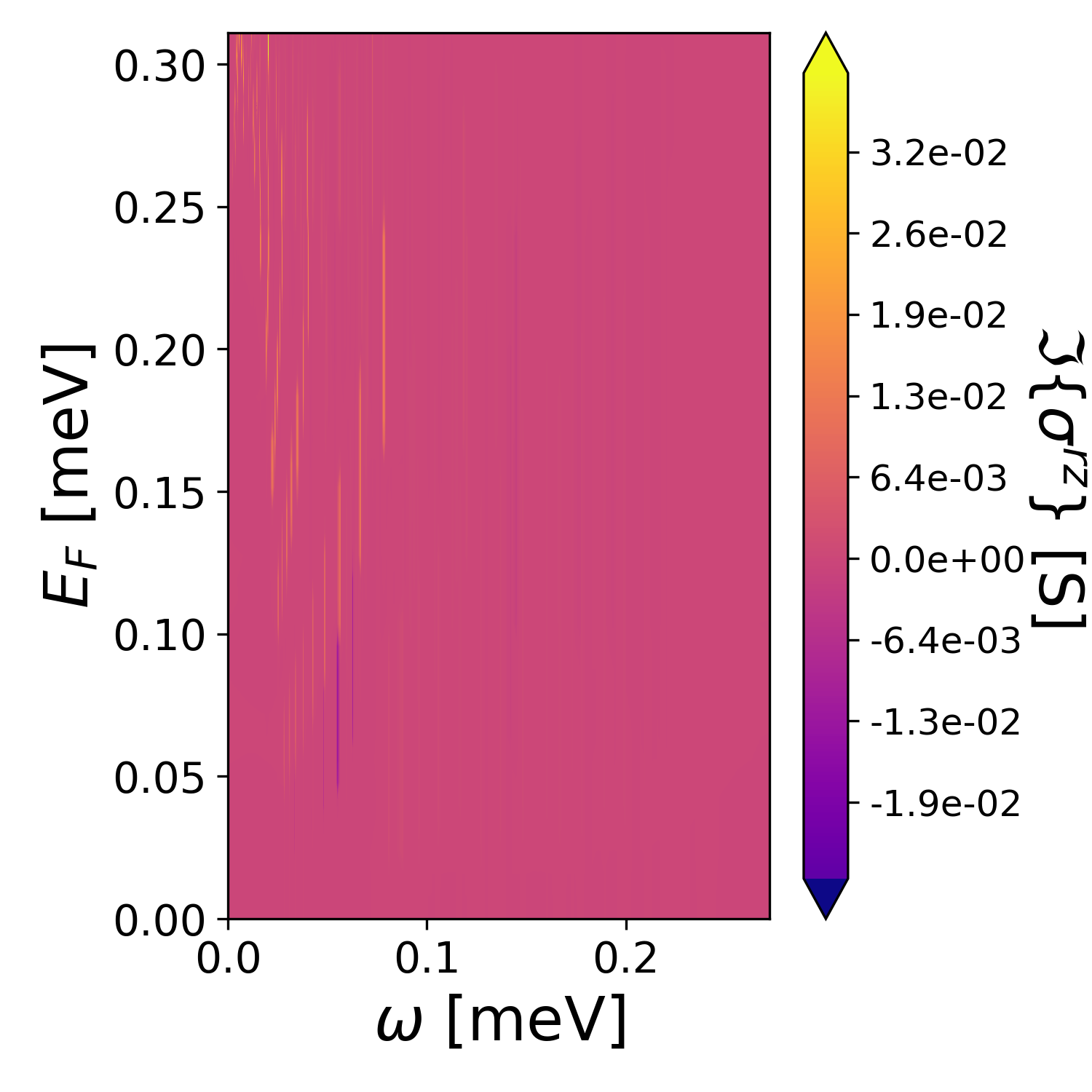}
         \caption{$g^*=1$}
     \end{subfigure}
     \hfill
     \begin{subfigure}[b]{0.3\textwidth}
         \centering
         \includegraphics[width=\textwidth]{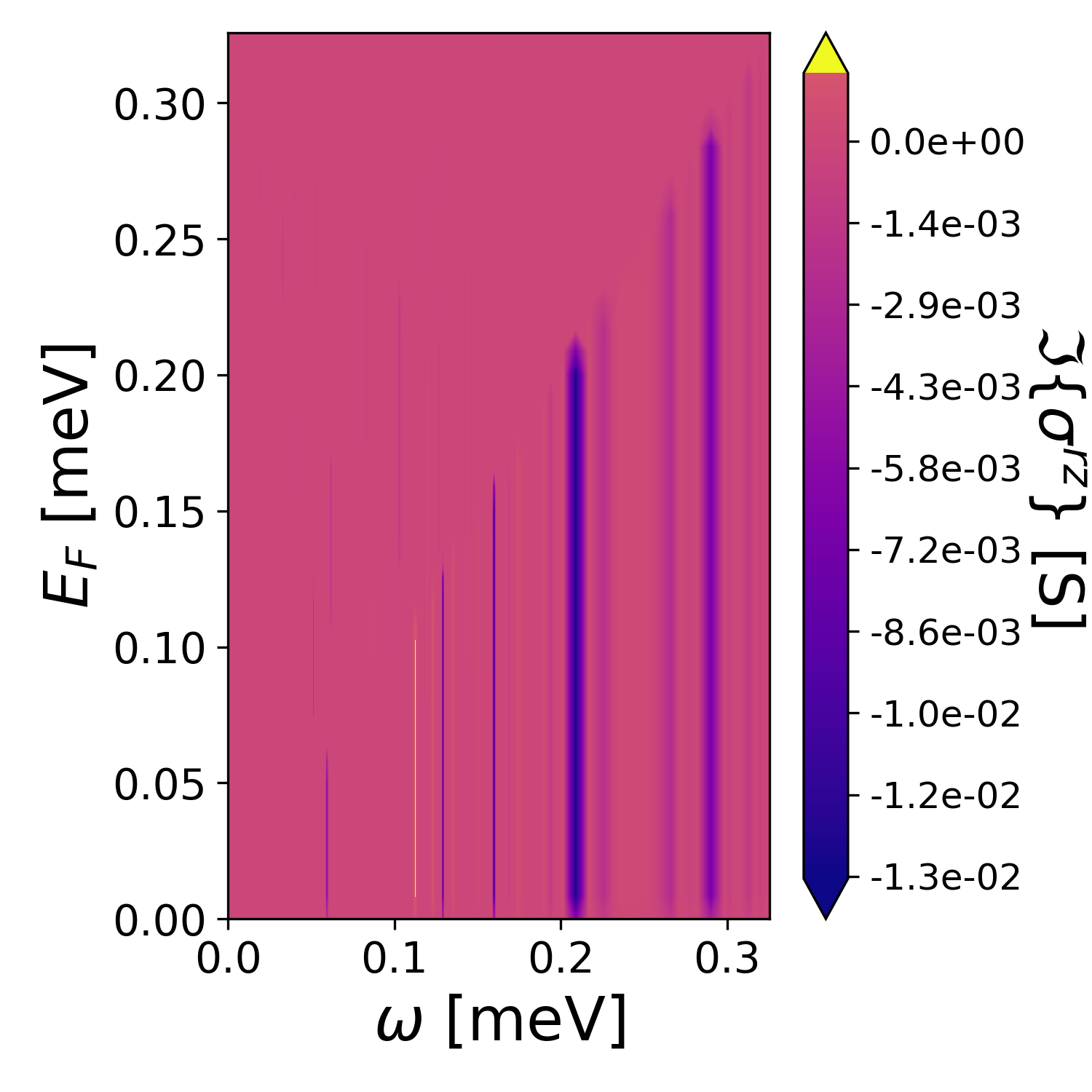}
         \caption{$g^*_c=2$}
     \end{subfigure}
     \hfill
     \begin{subfigure}[b]{0.3\textwidth}
         \centering
         \includegraphics[width=\textwidth]{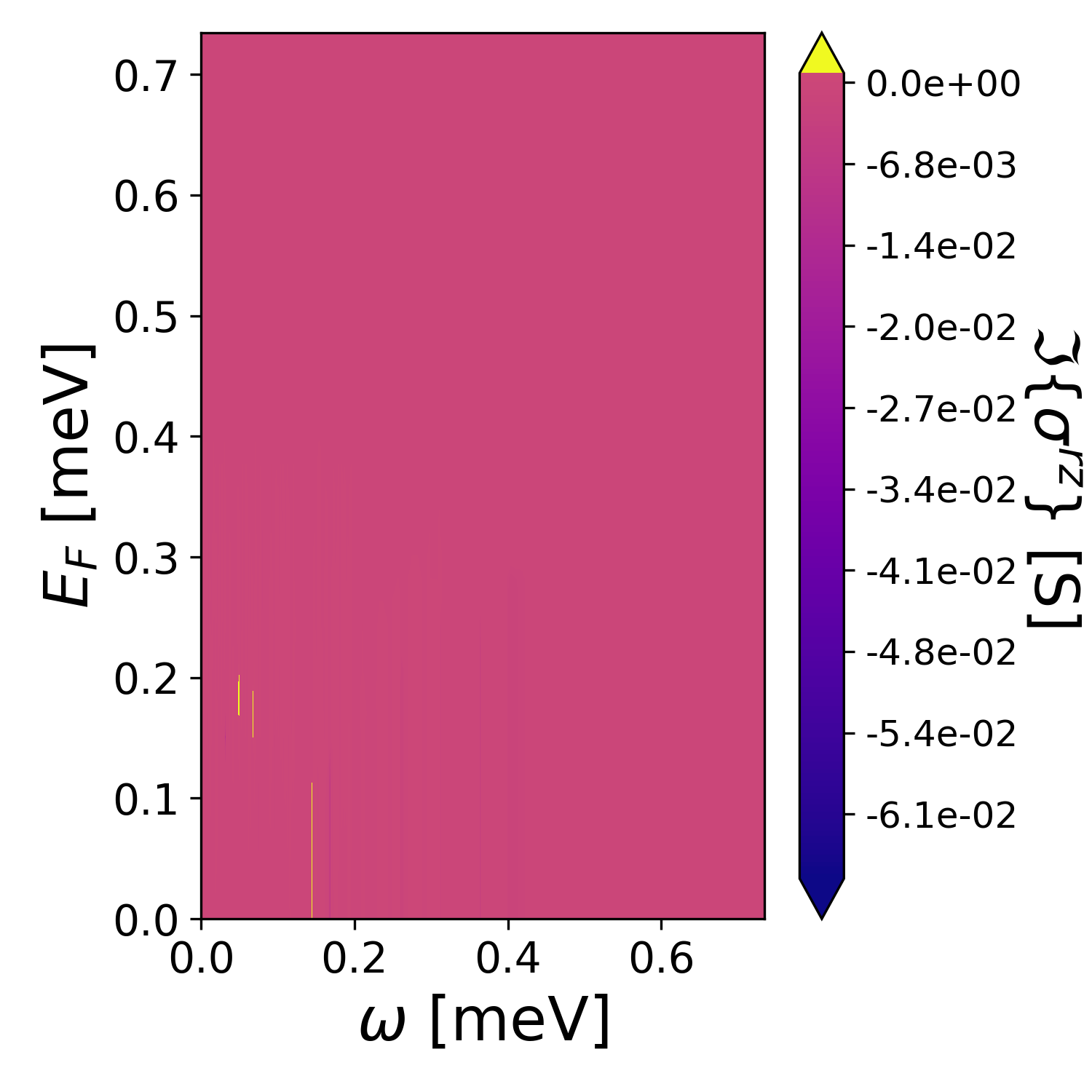}
         \caption{$g^*=3$}
     \end{subfigure}\\
        \caption{Absorptive radial spin-orbit Hall conductivity spectra with respect to different Fermi energies $E_F$ and driving frequencies $\omega$ of the external $E_z$-field perturbation. The shown spectral features for our InSb 2D electron gas are mainly caused by the local discretization of the 2D electron gas in the vicinity of the magnetic field divergence at the origin. In contrast, the more delocalized electronic states, that are further away from the origin, do not contribute significantly to the spectral features, which is a consequence of vanishing overlap matrix elements.  }
        \label{fig:spec}
\end{figure}

In a next step we compute the AC absorptive Hall conductivity under experimentally plausible conditions according to Eq. (\ref{eq:imac}).
For this purpose we focus on a 2D electron gas made of InSb, which possesses an extremely low effective electron mass of $0.015$ [a.u.] that makes it an ideal candidate to investigate spin-orbit effects in magnetic fields.\cite{lei2022high}. For our calculations we assume a temperature of $20$ [mK] and a static vector potential of $A_\phi =-6 \cdot10^{-4}$, which gives rise to a $1/r$ decaying magnetic field in z-direction on a mT-scale for radial distances from the origin in the $\mu m$-regime. The homogeneous perturbing electric field along the z-axis is chosen as $E_z = - 6\cdot 10^{-6}$ [a.u] which corresponds to $-3.1$ [V/$\mu$m].
Furthermore, we assume relatively long lived states by setting $\tau=1$ [ms] with corresponding sharp lineshape $\Gamma=1/\tau$ that partially compensates for the small prefactors of spin-orbit effects.
The numerical evaluation of Eq. (\ref{eq:imac}) simplifies considerably, when considering that the matrix element $\bra{a}\hat{d}_1\ket{b}$ imposes that only spin-flips can have a non-zero transition matrix elements, i.e.
\begin{eqnarray}
s_a=-s_b,\label{eq:spinsel}
\end{eqnarray}
 which leaves us with the spin-orbit current operator only, 
\begin{eqnarray}
\hat{j}^{r}=\hat{j}^{\rm SO,r}=-\frac{\hbar q}{4m^{* 2 }  c^2}\vec{\sigma}_j\times \vec{E}_z(0)\cdot \vec{e_r}=-\frac{\hbar q}{4m^{* 2} c^2 }(\sigma_y \cos\phi-\sigma_x\sin\phi)E_z.\label{eq:jrdef}
\end{eqnarray}
Eventually, Eqs. (\ref{eq:spinsel}) \& (\ref{eq:jrdef}) jointly determine the angular selection rule 
\begin{eqnarray}
l_a-l_b=\pm 1.\label{eq:orbsel}
\end{eqnarray}
Notice that those angular and spin selection rules remain valid for any radial symmetric magnetic fields. Furthermore, it also ensures that our restriction to bound states is valid, since any overlap with the (unknown) scattering states of the continuum must vanish except for the smallest angular momentum  quantum numbers $\min_l(l+g^* s/2)>0$, which separate the bound states from the continuum.\cite{sidler2022class} Eventually one can find a simplified expression for $\Im(\sigma^{rz}(\omega,E_F,T))$ as derived in App. \ref{app:conduct}, which was implemented for our numerical calculations. The resulting absorptive radial spin orbit Hall conductivity spectra is shown in Fig (\ref{fig:spec}) for different effective $g^*$-factors. Careful inspection of the AC spectrum reveals that the main contributions stem from localized electrons close to the origin, i.e., varying the finite number of numerical states does not change the strongest absorption peaks, which indicates that the overlap integrals become smaller, the more delocalized the eigenstates become. Thanks to this localisation effect we can converge our numerical calculations. This also suggests that the local / atomic nature of the AC absorption  is mainly related to the dacaying magnetic field and the infinite ranging vector potential plays a minor role. We find that for small $g^*=1$ significant absorption lines only occur for relatively low frequencies, whereas around $g^*=2$ a characteristic pattern occurs across the entire frequency regime. For the larger $g^*=3$ only a few significant absorption peaks occur. However, the range of allowed energetic transitions is significantly increased since the energetically lowest eigenvalue can become negative, while the ionization energy $E_{A^2}$ remains constant independently of $g^*$.

At this point, we would also like to mention two additional properties of the induced radial spin-orbit Hall current that are immediately evident from the analytics. First, the orientation of the perturbing field does not influence the radial current response, since
\begin{eqnarray}
j^r=-\sigma^{rz}(E_z,E_F,T) E_z\propto \frac{E_z^2}{m^{*4} c^4 }\label{eq:jpropto}
\end{eqnarray}
and furthermore it is suppressed by $m^{*4} c^4$, which suggests a tiny effect overall.
However, as we have seen in Fig. \ref{fig:spec},  a significant radial spin-orbit conductivity may still emerge for small effective electron masses and allowed transitions with long-lived states. Those aspects should be taken into account for a potential verification of the predicted effects.   

\section{Divergent absorption in static limit $\omega\rightarrow 0$}

\begin{figure}
     \begin{subfigure}[b]{0.3\textwidth}
         \centering
         \includegraphics[width=\textwidth]{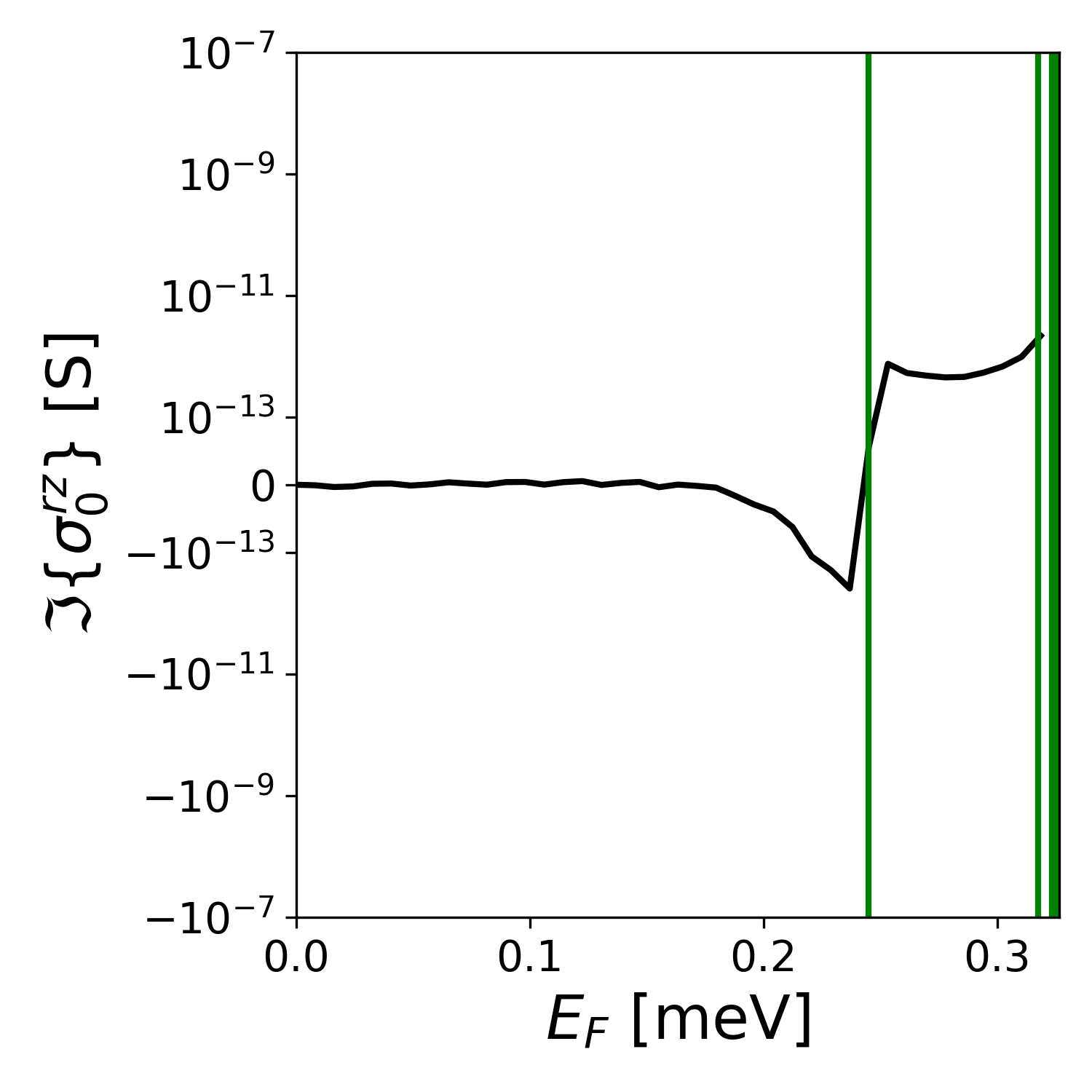}
         \caption{$g^*=1$}
         \label{fig:y equals x}
     \end{subfigure}
          \begin{subfigure}[b]{0.3\textwidth}
         \centering
         \includegraphics[width=\textwidth]{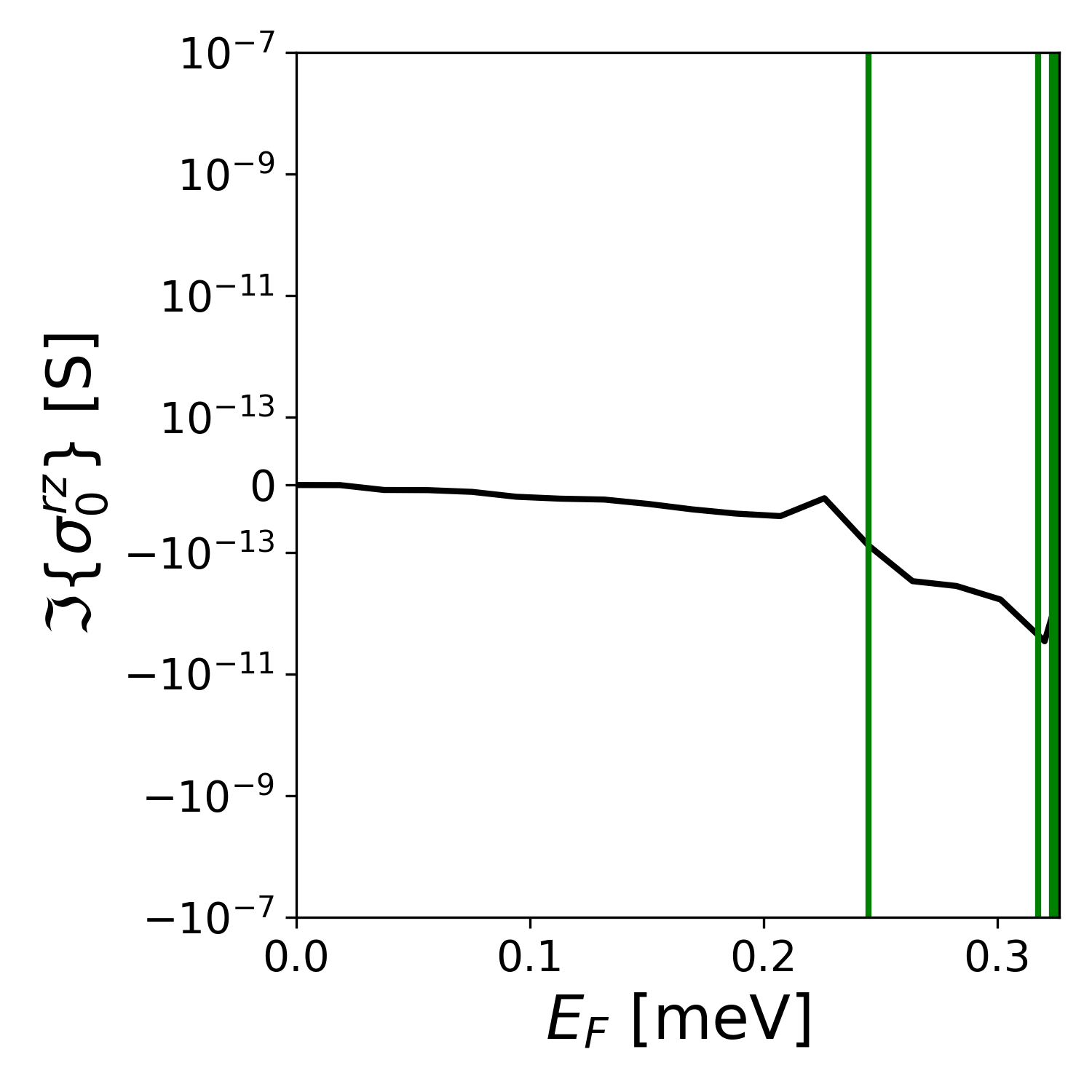}
         \caption{$g^*=3$}
         \label{fig:y equals x}
     \end{subfigure}\\
          \hfill
     \begin{subfigure}[b]{0.3\textwidth}
         \centering
         \includegraphics[width=\textwidth]{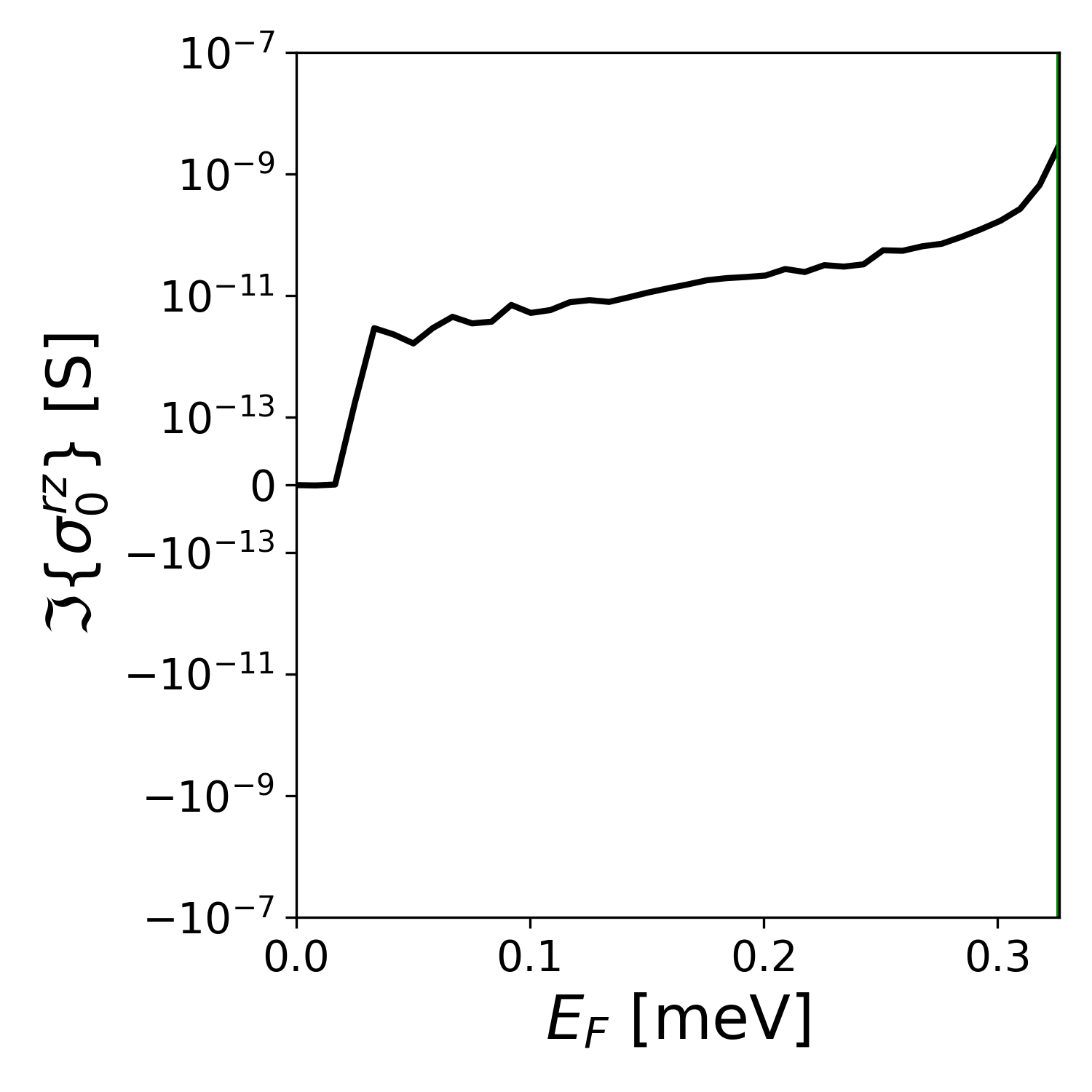}
         \caption{$g^*=4-g^*_{\rm rel}=1.9977$}
         \label{fig:three sin x}
     \end{subfigure}
     \hfill
     \begin{subfigure}[b]{0.3\textwidth}
         \centering
         \includegraphics[width=\textwidth]{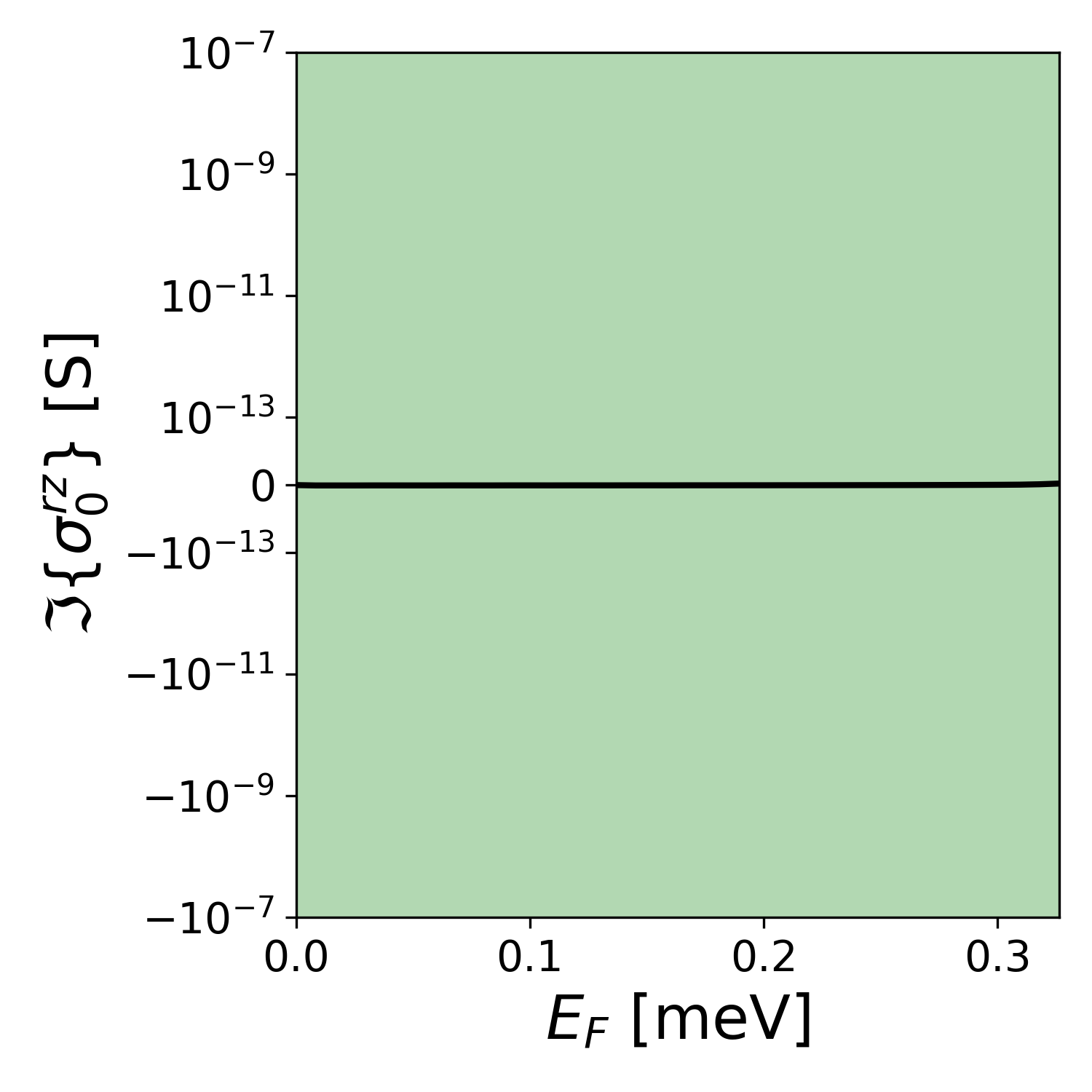}
         \caption{$g^*=2$}
         \label{fig:three sin x}
     \end{subfigure}
          \hfill
     \begin{subfigure}[b]{0.3\textwidth}
         \centering
         \includegraphics[width=\textwidth]{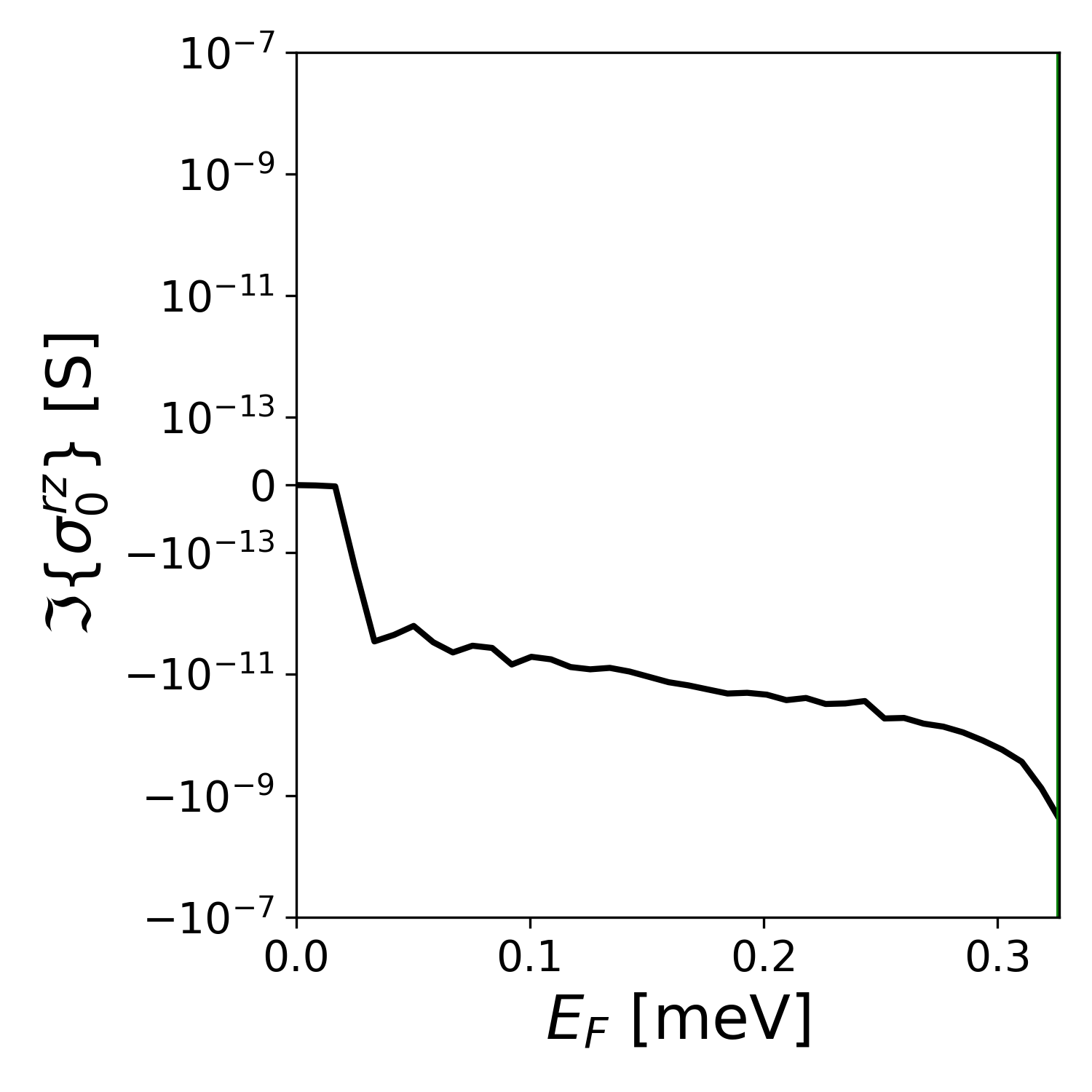}
         \caption{$g^*=g^*_{\rm rel}=2.0023$}
         \label{fig:three sin x}
     \end{subfigure}
        \caption{Absorptive (imaginary) part of the DC ($\omega\rightarrow0$) Hall conductivity with respect to the Fermi energy $E_F$ for different $g^*$-factors at $T=20$ [mK]. Quantum numbers were explored numerically up to $n=20$, which seem to yield converged results for Fermi energies not approaching the continuum limit at $E_F=E_{A^2}$. In this case only analytical considerations become feasible that indeed suggest a diverging absorptive Hall conductivity in the thermodynamic limit, i.e., for $n\rightarrow\infty$. }
        \label{fig:DCspec}
\end{figure}

In a next step, we  investigate the zero frequency limit $\Im(\sigma^{rz}_0(E_F,T))=\lim_{\omega\rightarrow0}\Im(\sigma^{rz}(\omega,E_F,T))$ of our setup, which shows particularly interesting features that originate from allowed transitions of degenerate states $E_a=E_b$. 
In the static limit Eq. (\ref{eq:imac}) reduces to, 
\begin{eqnarray}
\Im(\sigma^{rz}_0(E_F,T))
&=&\pi\iu \sum_{a,b}f(E_a)(1-f(E_b))\bigg[
\frac{\bra{a}\hat{j}^{r}\ket{b}\bra{b}\hat{d}_1 \ket{a}\Gamma}{(E_a-E_b)^2+\Gamma^2}-\frac{\bra{a}\hat{d}_1\ket{b}\bra{b}\hat{j}^{r} \ket{a}\Gamma}{ (E_b-E_a)^2+ \Gamma^2}\bigg].\label{eq:imdc},
\end{eqnarray}
which is evaluated for different $g^*$-factors and Fermi energies at $T=20$ [mK] of our InSb 2D electron gas in Fig. (\ref{fig:DCspec}).
Numerics reveals that overall the DC limit of the spin-orbit absorptive conductivity remains small, as one expects from Eq. (\ref{eq:jpropto}).  However, close inspection of $g^*$-factors in the vicinity of the critical non-relativistic $g^*_c:=2$ value reveal  a phase transition (i.e. sharp sign flip). In other words, by changing $g^*$, the external electric field will be  absorbed by either radially in-flowing or out-flowing currents. Similarly to the AC case, we reach numerical convergence only, with a finite number of states,  thanks to the localization of the significant overlap integrals. However, we notice that this is no longer true for Fermi energies approaching the ionization energy $E_{A^2}:=\lim_{n\rightarrow\infty} E_{n,l,s}=q^2A^2_\phi/(2m^*)$, where the conductivity starts to scale with the finite system size.  

In a next step, we try to better understand the aforementioned numerical absorption and scaling features  by analytical considerations in the zero temperature limit,
\begin{eqnarray}
\Im(\sigma^{rz}(\omega,E_F,0))
&=&\pi\iu \sum_{E_a<E_F\leq E_b}\bigg[
\frac{\bra{a}\hat{j}^{r}\ket{b}\bra{b}\hat{d}_1 \ket{a}\Gamma}{(E_a-E_b+\hbar\omega)^2+\Gamma^2}-\frac{\bra{a}\hat{d}_1\ket{b}\bra{b}\hat{j}^{r} \ket{a}\Gamma}{ (E_b-E_a+\hbar\omega)^2+ \Gamma^2}\bigg]\\
&=&-\pi\Gamma\frac{\hbar^2 q^2E_z}{16m^{*4}  c^4}\sum_{E_a<E_F \leq E_b}\bra{R_a}\ket{R_b}\cdot\label{eq:imT0a}\\
&&\begin{cases}
+\Big[\frac{\bra{R_b}
\frac{-\hbar l_a}{r}+qA_\phi- \hbar \partial_r \ket{R_a}}{(E_a-E_b+\hbar \omega)^2+\Gamma^2}+
\frac{\bra{R_a}\frac{-\hbar l_b}{r} +qA_\phi+ \hbar\partial_r\ket{R_b}}{(E_b-E_a+\hbar\omega)^2+ \Gamma^2}\Big]\delta_{l_a-l_b,1}\delta_{s_a,\downarrow}\delta_{s_b,\uparrow} & \text{for } 0<g^*<2 \\
       0 & \text{for } g^*=2\\
-\Big[\frac{\bra{R_b}
\frac{-\hbar l_a}{r}+qA_\phi +
\hbar \partial_r \ket{R_a}}{(E_a-E_b+\hbar \omega)^2+\Gamma^2}+
\frac{\bra{R_a}\frac{-\hbar l_b}{r} +qA_\phi- \hbar\partial_r\ket{R_b}}{(E_b-E_a+\hbar\omega)^2+ \Gamma^2}\Big]\delta_{l_a-l_b,-1}\delta_{s_a,\uparrow}\delta_{s_b,\downarrow}       & \text{for } g^*>2
        \end{cases}
         \nonumber,
\end{eqnarray}
where the different cases with respect to $g^*$ follow the selection rules of Eq. (\ref{eq:Aimac})  in combination with an exchange of the energetic order, i.e. $E_a\lesseqqgtr E_b\mapsto E_a\gtreqqless E_b$, when crossing the critical value $g^*_c=2$. While Eq. (\ref{eq:imT0a}) already suggests a phase transition, further insights can be achieved by investigating the transition matrix elements in the asymptotic limit at the ionization threshold, i.e. by taking $n\rightarrow\infty$ (i.e. energies closest to $E_{A^2}$)  
and considering states $E_a\overset{n\rightarrow\infty}\rightarrow E_b=E_F$ (see App. \ref{app:overlap}):
\begin{eqnarray}
    \bra{R_a}\ket{R_b}&=&\bra{R_b}\ket{R_a}\overset{n\rightarrow\infty}{=}-1 \label{eq:rarb}\\
    \bra{R_a}\frac{1}{r}\ket{R_b}&=&\bra{R_b}\frac{1}{r}\ket{R_a}\overset{n\rightarrow\infty}{=}0\\
    \lim_{n\rightarrow\infty}\bra{R_a}\partial_r\ket{R_b}&=&\lim_{n\rightarrow\infty}\bra{R_b}\partial_r\ket{R_a}=\frac{1}{2}.\label{eq:radrrb}
\end{eqnarray}
This suggests for the DC limit,
\begin{eqnarray}
|\lim_{n\rightarrow\infty}\Im(\sigma^{rz}_0(E_{F,n},0))|
&>&\left |-\pi\Gamma\frac{\hbar^2 q^3E_z A_\phi}{8m^{*4}  c^4}\sum_{E_a<E_F=E_b}\cdot\begin{cases}
\frac{1}{(E_a-E_b)^2+\Gamma^2}\delta_{l_a-l_b,1}\delta_{s_a,\downarrow}\delta_{s_b,\uparrow} & \text{for } 0<g^*<2 \\
       0 & \text{for } g^*=2\\
-\frac{1}{(E_a-E_b)^2+\Gamma^2}\delta_{l_a-l_b,-1}\delta_{s_a,\uparrow}\delta_{s_b,\downarrow}       & \text{for } g^*>2
        \end{cases}\right |,\nonumber\label{eq:imT0}
 \end{eqnarray}
 which implies
 \begin{eqnarray}
 \lim_{E_F\rightarrow E_{A^2}}\Im(\sigma^{rz}_0(E_F,0))     &=&-\pi\Gamma\frac{\hbar^2 q^3E_z A_\phi}{8m^{*4}  c^4}\left\{\begin{array}{lr}
\infty & \text{for } 0<g^*<2 \\
       0 & \text{for } g^*=2\\
-\infty   & \text{for } g^*>2
        \end{array}\right.\label{eq:divergence}.   
\end{eqnarray}
To proof that the summation in Eq. (\ref{eq:imT0}) indeed diverges in the thermodynamic limit, we need to find an infinite number of state pairs $\ket{a},\ket{b}$ that obey $E_a<E_F\leq E_b$. This can be seen by assuming $g^*=2+2\epsilon,\ \epsilon>0$  setting $E_F=E_b$ and solving $E_{n_a+k_c,l_a,1/2}= E_b=E_{n_a,l_a+1,-1/2}=E_F\overset{n_a\rightarrow\infty}{\rightarrow}E_{A}^2$ for integer $k_c$ which yields
\begin{eqnarray}
    k_c =\frac{\epsilon (2n_a+1)}{2l+1-\epsilon}\overset{n_a\rightarrow\infty}{\rightarrow} \infty,\ 0<\epsilon<2l+1,\label{eq:kc}
\end{eqnarray}
where an analogous  argument holds for the case $\epsilon<0$.
An unbounded $k_c$  implies that the summation in Eq. (\ref{eq:divergence}) indeed diverges, i.e. the number of contributing states grows at least with $k_c$, since they are either strictly positive or negative.  We consider the resulting divergent DC absorption by spin-orbit interaction  at the ionization threshold as one of the main results of this work. 
In other words,  Eq. (\ref{eq:divergence}) suggests that a (small) static, perpendicular electric field will be screened entirely by spin-orbit effects for Fermi energies approaching the ionization threshold! Depending on the $g^*$-factor, radially in- or out-flowing bound state currents will cause this effect, which can be re-interpreted as a longitudinal polarization effect by looking at the Maxwell equations. From $\vec{\nabla} \times\vec{B}=\mu_0(\vec{J}_b+\vec{J}_b+\epsilon_0 \partial_t \vec{E})$ where $\vec{J}_b$ indicates bound current density and $\vec{J}_f$ the free current density. The free currents can be safely discard for our setting as mentioned previously. We then notice that the curl of the magnetic field cannot account for the radial spin-orbit currents, which automatically implies that the spin-orbit current density $j_r(r)$ effectively correspond to a longitudinal polarization $P_r$ of the system in the macroscopic formulation of the Maxwell equations, i.e, $\partial_t P_r=j_r=-\sigma^{rz} E_z$. In other words, the divergent absorptive (imaginary) part of $\sigma^{rz}$ indicates that the external (perpendicular) electric field is entirely screened, as it would be the case inside (!) a perfect conductor.      
This screening effect emerges in our setup merely as a consequence of the constant vector potential $A_\phi$, i.e. it is an Aharonov-Bohm-like behaviour for infinite degeneracies in the thermodynamic limit.
Notice that the contributing states for $E_F\rightarrow E_{A^2}$ can be considered as radially  de-localized\cite{sidler2022class}. However,  the observed divergence is not only an effect of delocalization, but rather caused by the distorted Landau-level structure, which is a consequence of delicate interplay between the vector potential and the magnetic field (Zeeman interaction).\cite{sidler2022class} The resulting accumulation point at the ionization energy threshold can be occupied by any state with finite angular momentum in the thermodynamic limit (setting $n\rightarrow\infty$). Hence, this highly degenerate state of matter will at least theoretically be very sensitive towards all sorts of external DC perturbations. 
However, clearly, for any experimental realization things will be much more complex. For example, a perfect constant $A_\phi$-potential, i.e. a perfect $1/r$-magnetic field is not feasible, which in practice means that states far enough away from the origin will always belong to the continuum solution. In other words, finite system sizes will reduce the effect and complicate the experimental verification. Furthermore, beyond linear response effects may in principle lift the contributing energetic degeneracies. In addition, finite lifetimes and thermal effects for electronic fillings up to (almost) the continuum limit (at least for our ideal setup, not involving any band gaps) may also hamper an experimental realization. Nevertheless, we believe  distorted Landau levels should emerge in experiments with strong non-local behaviour, provided that on can realize a radial symmetric setup, with sufficiently localized magnetic field and long range vector potential.

Following this practical argument, it is tempting to ask if there are further divergences present in our system that the finite state numerics may have missed.
While we have observed divergent absorptive DC behaviour for the  fillings arbitrarily close to the ionization energy, the question emerges if a similar effect could also occur at lower Fermi energies. 
To shine further light on this question, we search for degenerate eigenvalues obeying the angular and spin selection rules as follows,   
%
\begin{eqnarray}
E_{n_a,l_a,1/2}(g^*)&=& E_{n_b,l_a+1,-1/2}(g^*). \label{eq:divergence_relation}
\end{eqnarray}
By  assuming a positive rational $g^*=g_1/g_2>0,\ \{g_1,g_2\}\in \mathbb{N}$ it turns out that we can find an infinite set of quantum numbers $n_a,l_a,n_b$ fulfilling Eq. (\ref{eq:divergence_relation}), given that there is at least one solution.
In more detail, we find a discrete set of infinitely degenerate energy levels at (see App. \ref{app:rydberg}),
\begin{eqnarray}
E^\infty(g^*\in \mathbb{Q}^*_+):=\begin{cases}
        E_{n,l,1/2}, n\neq l & \text{for } g^*=2 \\
        \frac{q^2 A_\phi^2}{2m^*}\Big(1-\Big[\frac{|2-g^*|}{2 k}\Big]^2\Big)>0,\  k\in \mathbb{O} & \text{otherwise. }
        \end{cases}\nonumber\label{eq:rydbergdivergence}
\end{eqnarray}
with $\mathbb{Q}^*_+$ being the strictly positive rational numbers and $\mathbb{O}$ the positive odd integer numbers.
The resulting energy pattern  resembles a squeezed Rydberg-like series that strongly depends on $g^*$, as can be seen by the vertical green lines in Fig. \ref{fig:DCspec}. Nevertheless, they all share an accumulation point at $E^\infty \rightarrow E_{A^2}$, independently of $g^*$,  as can be seen for $k\rightarrow\infty$, which is in line with the previous considerations at the ionization threshold. Notice also that the  density of the degenerate bands will in practice delicately depend on the chosen $g^*$-factor. For our ideal $1/r$-magnetic field setup, numerics suggests that the Rydberg-like bands contribute only locally to the DC conductivity $\Im(\sigma^{rz}_0(E_F,0))$ as it was the case for the AC response, since the matrix overlap seems to strongly decline for states localized further away from the origin. This drastically limits the absorbance for Fermi energies close to $E^\infty(g^*)$ except for the previously discussed divergent case at the ionization threshold. However, this strong local suppression may no longer hold for different radial symmetric field shapes, which in turn could facilitate the detection of strong spin-orbit effects at specific Fermi energies,  similarly to our ideal system at the ionization threshold.

\section{Analytic solution and protected degeneracies for time-dependent magnetic fields}

\begin{figure}
     \begin{subfigure}[b]{0.6\textwidth}
         \centering
         \includegraphics[width=\textwidth]{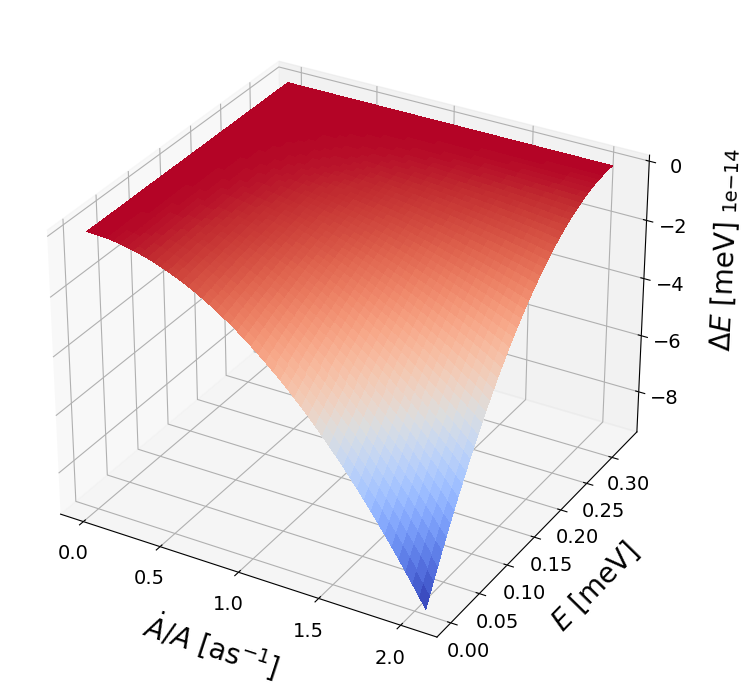}
     \end{subfigure}
        \caption{Deviation of the time-dependent energy eigenvalue $\tilde{E}$ for the chosen  InSb setup, with respect to the strictly static energy $E$ measured by $\Delta E:= \tilde{E}(E)-E$ for different small variations of the externally applied vector potential $A_\phi$. Notice, the extremely small shift due to non-zero $\dot{A}_\phi$. However, as explained in the main text, the much more striking feature is that the structure of the static energy eigenvalues remains preserved exactly, i.e. $\tilde{E}(E)$. }
        \label{fig:dE}
\end{figure}

We have seen so far that our system possesses  intriguing linear response features that emerge in the DC limit due to the presence of (infinitely) degenerate states. This immediately rises the question of how robust those degeneracies are. As already mentioned, they will overall be very delicate to achieve in experiments for multiple reasons. However, surprisingly they seem to be protected with respect to slow enough changes of the applied strength of the $1/r$-magnetic field, as we subsequently will show.
It turns out that we can find an analytical solutions for the non-relativistic Hamiltonian expanded up to order $1/(m^*c)^2$,\cite{frohlich1993gauge} i.e. including the spin-orbit and Darwin interaction, in a similar spirit to Ref. \citenum{sidler2022class}. In more detail we can solve the eigenvalue problem for, 
\begin{eqnarray}
\hat{\tilde{H}}(t)&=&\sum_{j=1}^N\frac{1}{2m^*}\hat{\vec{\Pi}}_j^2-\frac{g^* q\hbar}{4 m^*} \vec{\sigma}_j\cdot\vec{B}(\vec{r}_j,t)\nonumber\\
&&-\frac{q\hbar}{8m^{*2}c^2}\Big[\hat{\vec{\Pi}}_j\cdot\big(\vec{\sigma}_j\times\vec{E}(\boldsymbol{r}_j,t)\big)+\big(\vec{\sigma}_j\times\vec{E}(\boldsymbol{r}_j,t)\big)\cdot\hat{\vec{\Pi}}_j\Big]\label{eq:hamiltonianmc2}\\
&&-\frac{q\hbar}{8m^{*2} c^2} \mathrm{div}\vec{E}(\boldsymbol{r}_j,t),\nonumber
\end{eqnarray}
as well, instead of only solving Eq. (\ref{eq:hamiltonian0}). We now assume slowly varying  external fields of the following form $\vec{A}(\vec{r},t):=A_\phi(t)\vec{e}_\phi$ and $\vec{B}(\vec{r},t)=A_\phi(t)/r \vec{e}_z$. The
corresponding electric field is given by $\vec{E}(\boldsymbol{r},t)=-\nabla_{\boldsymbol{r}} V/e-(\partial/\partial_t)\vec{A} = -\dot{A}_\phi(t) \vec{e}_\phi $, which immediately removes the Darwin interaction (last term) in our Hamiltonian formulation.  By applying the method of Frobenius, we can determine the bound state spectrum (see App. \ref{app:analytic_solution}),

\begin{eqnarray}
\tilde{E}_{n,l,s}(t)=\frac{q^2 A_\phi^2
}{2m^*}-\frac{16m^{*3}  c^4}{q^2 \hbar^2 \dot{A}_\phi^2}\Bigg[1-\sqrt{1-\frac{q^4 \hbar^2 \dot{A}_\phi^2A_\phi^2
}{16m^{*4}  c^4}\bigg[\frac{2l+g^* s}{2n+1}\bigg]^2}\Bigg],\ n\geq l,\ l+g^* s/2>0.\label{eq:etilde}
\end{eqnarray}
While on a first glance the spectrum looks much different compared to the static Eq. (\ref{eq:Etotbound}), by taking the static limit one indeed recovers
\begin{eqnarray}
\lim_{\dot{A}\rightarrow 0}\tilde{E}_{n,l,s}(t)&=&E_{n,l,s},
\end{eqnarray}
as one would expect.
More remarkable, however, is that the spin-orbit interaction, which now contributes due to the time-dependent vector potential, does not change the structure of the spectrum, i.e., degeneracies are not lifted, since one can reformulate Eq. (\ref{eq:etilde}) as follows,
\begin{eqnarray}
\tilde{E}(E_{n,l,s}(t),t)&=&\frac{q^2 A_\phi^2
}{2m^*}-\frac{16m^{*3}  c^4}{q^2 \hbar^2 \dot{A}_\phi^2}\Bigg[1-\sqrt{1+\frac{q^2 \hbar^2 \dot{A}_\phi^2
}{8m^{*3} c^4 }(E_{n,l,s}(t)-E_{A^2})}\Bigg].\label{eq:etilde_e}
\end{eqnarray}
Consequently, the degenerate states seem to be protected with respect to slow variations of $A_\phi$ at least up to order $1/(m^* c)^2$. However, the overall  squeezing of the energetic structure is extremely small as one can see in   Fig. \ref{fig:dE} for moderate time-variations $\dot{A}$. Clearly, this picture might change for jumps in the magnetic field, but then one automatically violates the  assumption of a slow field variation, i.e. one would need to solve the time-dependent Schroedinger problem instead.

 We can not only determine analytic eigenvalues, but also the corresponding orthonormal eigenstates in terms of generalized Laguerre polynomials $L^{\nu}_w(x)$, see App. (\ref{app:analytic_solution}),
\begin{eqnarray}
\tilde{\Psi}_{n,l,s}(t)=\frac{1}{\sqrt{\tilde{N}_{n,l,s}}}e^{il\phi}e^{-\iu \gamma(r)}e^{-\frac{\tilde{x}(r)}{2}}\tilde{x}^l(r) L_{n-l}^{2l}(\tilde{x}(r))\chi(s),\label{eq:tildeeigf}
\end{eqnarray}
which are identical to the static solution given in Eq. (\ref{eq:eigf}), except for a $r$-dependent complex phase factor, 
\begin{eqnarray}
    \gamma(r):= \frac{\delta s}{\sqrt{1-\delta^2}} x(r)
\end{eqnarray}
and a different scaling in radial coordinates, 
\begin{eqnarray}
\tilde{x}(r):=\sqrt{1-\delta^2}\frac{2 q A_\phi}{\hbar}\frac{2l+g^* s}{2n+1}r,\label{eq:scalingtilde}
\end{eqnarray}
where $\delta:=\frac{q\hbar^2}{4m^{*2}c^2}\dot{A}_\phi \sqrt{\frac{m^* |\tilde{E}^-|}{2\hbar^2}}$ with $\tilde{E}^-:=\tilde{E}-E_{A^2}$.
The corresponding normalisation is explicitly calculated as,
\begin{eqnarray}
\tilde{N}_{n,l,s}&=&\int_0^{2\pi}\int_0^\infty \tilde{\Psi}_{n,l,s}^*\tilde{\Psi}_{n,l,s} rdr d\phi\nonumber\\
&=&2\pi \frac{(n+l)!}{(n-l)!}(2n+1)\bigg(\sqrt{1-\delta^2}\frac{2 q A_\phi}{\hbar}\frac{2l+g^* s}{2n+1}\bigg)^{-2}.\label{eq:tilde_norm}
\end{eqnarray}
Notice that for constant vector potential $\dot{A}_\phi=0$, we naturally recover the previously known static solution as given in Eq. (\ref{eq:eigf}). 
The emergence of a radial dependent phase factor opens interesting research perspectives with regards to geometric phases and topological protection for slow variations of $A_\phi,\dot{A}_\phi$, which may substantiate theoretically the protected energetic structure of Eq. (\ref{eq:etilde_e}). However, this clearly goes beyond the scope of this work.

\section{Conclusion and outlook}

We have  investigated the effect of spin-orbit interaction in a 2D electron gas with distorted Landau level structure that arises from an externally applied radial symmetric, $1/r$-decaying magnetic field. By applying a weak perpendicular electric field  $\vec{E}_z$  with respect to the 2D electron gas (and in parallel with the applied magnetic field), we find that radial Hall currents emerge solely from spin-orbit interaction. The absorptive part of the radial Hall conductivity shows spectral features that are of atomistic  nature, i.e., dominated by the discrete energy levels localized close to the origin. While overall the magnitude of absorption is expected to be small, because of the $1/(m^{*}c)^4$-prefactor and the quadratic dependency of the radial Hall current on the $\vec{E}_z$ perturbation. Partially, this suppression can  be mediated by choosing a small effective electron mass of the host material (InSb) and by assuming long-lived states.  
Surprisingly, things change when approaching Fermi energies close to the ionization threshold, where we could show analytically that a divergent static spin-orbit absorbance emerges in the thermodynamic limit due to the distorted Landau level structure.  This divergent linear response implies that the external electric field is entirely absorbed by induced radial spin-orbit currents, i.e. by radial polarization, where the induced flow / polarization direction  depends on the effective $g^*$-factor of the material. In other words, the small electric field perpendicular to the 2D electron gas is screened entirely, as it would be the case inside a perfect conductor. Interestingly, the divergent bound state currents also show a  critical behaviour at $g^*_c=2$, where the radial current response is reversed.
In our setup, the physical origin of the divergence  is a consequence of the long-range (const) vector potential (Aharonov-Bohm-like effect), which clearly in this form is experimentally out of reach, due to the associated magnetic field divergence at the origin. 
However, for other radial magnetic field, which induce differently distorted Landau levels, we would still  expect the generic emergence of some discrete level structure with possibly strong linear response to external perturbations. Probably in a similar fashion to our case, for which we found a discrete, infinitely degenerate Rydberg-like band structure that obeys allowed angular momentum and spin selection rules. While this structure turns out to be of minor relevance for our setup, things may change for different (more realistic) radial field dependencies.
On should also keep in mind that the proposed spin-orbit effects may depend non-trivially on many other external influences (e.g. Coulomb interaction, decoherence, impurities). Nevertheless, we could show analytically  that the structure of the energy eigenvalues remains preserved exactly with respect to slow variations of the applied $1/r$-magnetic field strength, by solving the Dirac equation expanded up to order $1/(m^*c)^2$. In particular, we find that the time-dependent field variations cannot lift the static degeneracies, which suggests a protection mechanism that will become the focus of future  work.
Clearly, there is still a long way to go, with many obstacles to overcome, for an experimental verification of our theoretical predictions. However, we believe the experimental realization of  rotationally distorted Landau levels would not only be highly fruitful with respect to spin-orbit effects, but more generally by opening novel pathways to engineer (distorted) Landau level physics from local to non-local degeneracies.

\begin{acknowledgments}
We thank Jerome Faist and Gregor Jotzu for inspiring discussions and helpful comments.
This work was made possible through the support of the RouTe Project (13N14839), financed by the Federal Ministry of Education and Research (Bundesministerium für Bildung und Forschung (BMBF)) and supported by the European Research Council (ERC-2015-AdG694097), the Cluster of Excellence “CUI: Advanced Imaging of Matter” of the Deutsche Forschungsgemeinschaft (DFG), EXC 2056, project ID 390715994 and the Grupos Consolidados (IT1249-19).
The Flatiron Institute is a division of the Simons Foundation.
Author contributions: D.S. initiated the project, derived the analytic solutions, performed numerical calculations and discovered the spin-orbit absorption divergence. M.R. contributed to the mathematical accuracy and rigorousness. All authors developed the physical interpretation and wrote the manuscript.
\end{acknowledgments}

\section*{Data Availability Statement}
Python implementation available on reasonable request.

\bibliographystyle{unsrt}
\bibliography{manuscript}
\newpage
\appendix
\section{Radial spin-orbit Hall conductivity simplification \label{app:conduct}}
For our radial symmetric $1/r$-decaying magnetic field setup the radial spin-orbit hall conductivity can be simplified by using the angular and spin-orbit selection rules that enter as follows,
%
\begin{eqnarray}
\bra{s_a,l_a}\sigma_y \cos\phi-\sigma_x\sin\phi\ket{s_b,l_b}&=&\begin{cases}
        -\iu, & \text{for }l_a-l_b=-1,s_a=\uparrow,s_b=\downarrow\\
        \iu, & \text{for }l_a-l_b=1,s_a=\downarrow,s_b=\uparrow\\
        0, & \text{otherwise}
        \end{cases}\\
\bra{s_a,l_a}-\sigma_y \sin\phi-\sigma_x\cos\phi\ket{s_b,l_b}&=&\begin{cases}
        -1, & \text{for }l_a-l_b=-1,s_a=\uparrow,s_b=\downarrow\\
        -1, & \text{for }l_a-l_b=1,s_a=\downarrow,s_b=\uparrow\\
        0, & \text{otherwise}
        \end{cases}\\\nonumber
\end{eqnarray}
which eventually yields,
\begin{eqnarray}
\Im(\sigma^{rz}(\omega,E_F,T))
&=&\pi\iu \sum_{a,b}f(E_a)(1-f(E_b))\bigg[
\frac{\bra{a}\hat{j}^{r}\ket{b}\bra{b}\hat{d}_1 \ket{a}\Gamma}{(E_a-E_b+\hbar \omega)^2+\Gamma^2}-\frac{\bra{a}\hat{d}_1\ket{b}\bra{b}\hat{j}^{r} \ket{a}\Gamma}{ (E_b-E_a+\hbar\omega)^2+ \Gamma^2}\bigg]\\
&=&\pi\iu \Gamma\frac{\hbar q}{4m^{*2}  c^2}\sum_{a,b}f(E_a)(1-f(E_b))\bra{a}\sigma_y \cos\phi-\sigma_x\sin\phi\ket{b}\\
&&\bigg[
\frac{\bra{b}\hat{d}_1 \ket{a}}{(E_a-E_b+\hbar \omega)^2+\Gamma^2}+\frac{\bra{a}\hat{d}_1\ket{b}}{ (E_b-E_a+\hbar\omega)^2+ \Gamma^2}\bigg]\nonumber\\
&=&\pi\iu \Gamma\frac{\hbar^2 q^2E_z}{16m^{*4}  c^4}\sum_{a,b}f(E_a)(1-f(E_b))\bra{a}\sigma_y \cos\phi-\sigma_x\sin\phi\ket{b}\\
&&\Big[\frac{\bra{b}(-\sigma_y \sin\phi-\sigma_x\cos\phi)
\cdot(-\frac{\iu \hbar}{r} \partial_\phi-qA_\phi) +(\sigma_y \cos\phi-\sigma_x\sin\phi)
\cdot(-\iu \hbar \partial_r) \ket{a}}{(E_a-E_b+\hbar \omega)^2+\Gamma^2}+\nonumber\\
&&\frac{\bra{a}(-\sigma_y \sin\phi-\sigma_x\cos\phi)(-\frac{\iu \hbar}{r} \partial_\phi-qA_\phi)+
(\sigma_y \cos\phi-\sigma_x\sin\phi)(-\iu \hbar\partial_r)\ket{b}}{(E_b-E_a+\hbar\omega)^2+ \Gamma^2}\Big]\nonumber\\
&=&\pi\iu \Gamma\frac{\hbar^2 q^2E_z}{16m^{*4}  c^4}\sum_{a,b}f(E_a)(1-f(E_b))\bigg\{\\
&&-i\bra{R_a}\ket{R_b}\Big[\frac{\bra{R_b}(-1)
\cdot(\frac{\hbar l_a}{r}-qA_\phi) +(i)
\cdot(-\iu \hbar \partial_r) \ket{R_a}}{(E_a-E_b+\hbar \omega)^2+\Gamma^2}+\nonumber\\
&&\frac{\bra{R_a}(-1)(\frac{\hbar l_b}{r} -qA_\phi)+(-\iu)(-\iu \hbar\partial_r)\ket{R_b}}{(E_b-E_a+\hbar\omega)^2+ \Gamma^2}\Big]\delta_{l_a-l_b,-1}\delta_{s_a,\uparrow}\delta_{s_b,\downarrow}\nonumber\\
&&+i\bra{R_a}\ket{R_b}\Big[\frac{\bra{R_b}(-1)
\cdot(\frac{\hbar l_a}{r}-qA_\phi) +(-i)
\cdot(-\iu \hbar \partial_r) \ket{R_a}}{(E_a-E_b+\hbar \omega)^2+\Gamma^2}+\nonumber\\
&&\frac{\bra{R_a}(-1)(\frac{\hbar l_b}{r} -qA_\phi)+(\iu)(-\iu \hbar\partial_r)\ket{R_b}}{(E_b-E_a+\hbar\omega)^2+ \Gamma^2}\Big]\delta_{l_a-l_b,1}\delta_{s_a,\downarrow}\delta_{s_b,\uparrow}\bigg\}\nonumber\\
%
%
&=&-\pi\Gamma\frac{\hbar^2 q^2E_z}{16m^{*4}  c^4}\sum_{a,b}f(E_a)(1-f(E_b))\bigg\{ \label{eq:Aimac}\\
&&-\bra{R_a}\ket{R_b}\Big[\frac{\bra{R_b}
\frac{-\hbar l_a}{r}+qA_\phi +
\hbar \partial_r \ket{R_a}}{(E_a-E_b+\hbar \omega)^2+\Gamma^2}+\nonumber\\
&&\frac{\bra{R_a}\frac{-\hbar l_b}{r} +qA_\phi- \hbar\partial_r\ket{R_b}}{(E_b-E_a+\hbar\omega)^2+ \Gamma^2}\Big]\delta_{l_a-l_b,-1}\delta_{s_a,\uparrow}\delta_{s_b,\downarrow}\nonumber\\
&&+\bra{R_a}\ket{R_b}\Big[\frac{\bra{R_b}
\frac{-\hbar l_a}{r}+qA_\phi- \hbar \partial_r \ket{R_a}}{(E_a-E_b+\hbar \omega)^2+\Gamma^2}+\nonumber\\
&&\frac{\bra{R_a}\frac{-\hbar l_b}{r} +qA_\phi+ \hbar\partial_r\ket{R_b}}{(E_b-E_a+\hbar\omega)^2+ \Gamma^2}\Big]\delta_{l_a-l_b,1}\delta_{s_a,\downarrow}\delta_{s_b,\uparrow}\bigg\}.\nonumber
\end{eqnarray}


\section{Radial overlap integrals \label{app:overlap}}
For analytical accessiblity, we assume that $E_a\overset{n\rightarrow\infty}{\rightarrow}E_b$, which means that the radial scaling factor $x(r)=:\nu r$, defined from Eq. (\ref{eq:scaling}), becomes equal for both states $\ket{a}$ and $\ket{b}$. Notice that $\nu\propto 1/(2n+1)\overset{n\rightarrow\infty}{\rightarrow} 0$. The resulting radial overlap integrals can then be solved analytically
\begin{eqnarray}
    \bra{R_a}\ket{R_b}&\overset{\nu_a=\nu_b}{=}&\frac{1}{\sqrt{N_l N_{l+1}}} \int_0^\infty e^{-\nu r} (\nu r)^{2l+1} L_{n-l}^{2l}(\nu r)L_{n-l-1}^{2l+2}(\nu r) r dr\\
    &=&\frac{1}{\sqrt{N_l N_{l+1}}\nu^2} \int_0^\infty e^{-\rho} (\rho)^{2l+2} L_{n-l}^{2l}(\rho)L_{n-l-1}^{2l+2}(\rho)  d\rho\\
    &=&\frac{1}{\sqrt{N_l N_{l+1}}\nu^2}  \int_0^\infty e^{-\rho} (\rho)^{2l+2} \big[L_{n-l}^{2l+2}(\rho)-2L_{n-l-1}^{2l+2}(\rho)+L_{n-l-2}^{2l+2}(\rho)\big]\cdot\\
    &&L_{n-l-1}^{2l+2}(\rho)  d\rho\nonumber\\
    &=&-2\sqrt{\frac{(n-l)!}{(n+l)!(2n+1)}\frac{(n-l-1)!}{(n+l+1)!(2n+1)}}  \frac{(n+l+1)!}{(n-l-1)!}\\
    &\overset{n\rightarrow\infty}{=}& -1,
\end{eqnarray}
where we have used the following well known 3-point rule and orthogonality relation for the generalized Laguerre polynomials,\cite{abramowitz1988handbook}
\begin{eqnarray}
    L_n^\alpha& =&L_n^{\alpha+1}-L_{n-1}^{\alpha+1}\\
    \int_0^\infty x^\alpha e^{-\alpha}L_n^\alpha(x)L_m^\alpha(x) dx &=& \frac{\Gamma(n+\alpha+1)}{n!}\delta_{n,m}.
\end{eqnarray}
with,\cite{sidler2022class}
\begin{eqnarray}
    N_l&:=& \nu^{-2}\frac{(n+l)!}{(n-l)!}(2n+1),
\end{eqnarray}
and Mathematica \cite{Mathematica} for the limiting procedure.

In a next step we similarly investigate,
\begin{eqnarray}
    \bra{R_a}\frac{1}{r}\ket{R_b}&\overset{\nu_a=\nu_b}{=}&\frac{1}{\sqrt{N_l N_{l+1}}} \int_0^\infty e^{-\nu r} (\nu r)^{2l+1} L_{n-l}^{2l}(\nu r)L_{n-l-1}^{2l+2}(\nu r)  dr\\
    &=&\frac{1}{\sqrt{N_l N_{l+1}}\nu} \int_0^\infty e^{-\rho} (\rho)^{2l+1} L_{n-l}^{2l}(\rho)L_{n-l-1}^{2l+2}(\rho)  d\rho\\
    &=&\frac{1}{\sqrt{N_l N_{l+1}}\nu}  \int_0^\infty e^{-\rho} (\rho)^{2l+1} \big[L_{n-l}^{2l+1}(\rho)-L_{n-l-1}^{2l+1}(\rho)\big]\\
    &&\big[L_{n-l-1}^{2l+1}(\rho) +L_{n-l-2}^{2l+1}(\rho)+...+L_{1}^{2l+1}(\rho)+1\big]  d\rho\nonumber\\
    &=&\nu\sqrt{\frac{(n-l)!}{(n+l)!(2n+1)}\frac{(n-l-1)!}{(n+l+1)!(2n+1)}}\cdot\\ 
    &&\bigg(\frac{(n+l+1)!}{(n-l)!}-\frac{(n+l)!}{(n-l-1)!}\bigg)\nonumber\\
    &\overset{n\rightarrow\infty}{=}&0,
\end{eqnarray}
and we find,
\begin{eqnarray}
    \bra{R_a}\partial_r\ket{R_b}&\overset{\nu_a=\nu_b}{=}&\frac{1}{\sqrt{N_l N_{l+1}}} \int_0^\infty e^{-\frac{\nu r}{2}} (\nu r)^{l+1} L_{n-l-1}^{2l+2}(\nu r)\big[\partial_r e^{-\frac{\nu r}{2}} (\nu r)^l L_{n-l}^{2l}(\nu r)\big] r dr\\
&=&- \bra{R_a}\ket{R_b}+l \bra{R_a}\frac{1}{r}\ket{R_b}-\frac{1}{\sqrt{N_l N_{l+1}}} \int_0^\infty e^{-\nu r} (\nu r)^{2l+1} L_{n-l-1}^{2l+1}(\nu r)L_{n-l-1}^{2l+2}(\nu r) r dr\\
&=&- \bra{R_a}\ket{R_b}+l \bra{R_a}\frac{1}{r}\ket{R_b}\\
&&-\frac{1}{\sqrt{N_l N_{l+1}}\nu^2}  \int_0^\infty e^{-\rho} (\rho)^{2l+2} \big[L_{n-l-1}^{2l+2}(\rho)-L_{n-l-2}^{2l+2}(\rho)\big]L_{n-l-1}^{2l+2}(\rho)  d\rho\nonumber\\
&=&- \bra{R_a}\ket{R_b}+l \bra{R_a}\frac{1}{r}\ket{R_b}\\
&&-\sqrt{\frac{(n-l)!}{(n+l)!(2n+1)}\frac{(n-l-1)!}{(n+l+1)!(2n+1)}}  \frac{(n+l+1)!}{(n-l-1)!}\nonumber\\
&\overset{n\rightarrow\infty}{=}&\frac{1}{2}.
\end{eqnarray}
Analogous  derivations can be applied to show  that $\bra{R_b}\partial_r\ket{R_a}\rightarrow1/2$.

\section{Derivation squeezed Rydberg divergences \label{app:rydberg}}

To solve the integer value problem imposed by the energy degeneracy relation given in Eq. (\ref{eq:divergence_relation}), three different regimes of rational $g^*:=g_1/g_2>0$ have to be considered separately. For $0<g^*<2$ we find,
\begin{eqnarray}
\frac{2l_1+g_1/(2g_2)}{2n_1+1}&=&\frac{2l_1+g_1/(2g_2)}{2n_2+1}+\frac{2(1-g_1/(2g_2))}{2n_2+1}\overset{!}{=}C\\
n_2&:=&n_1+\Delta n, \Delta n\in \mathbb{N}\ \mathrm{with}\ \Delta n\leq n_1\\
2g_2(2(n_1+\Delta n)+1)\frac{2l_1+g_1/(2g_2)}{2n_1+1}&=&4g_2l_1+g_1+2(2g_2-g_1)\\
&\Rightarrow&\nonumber\\
\Delta n&=&\frac{(2g_2-g_1)(2n_1+1)}{4g_2 l_1+g_1}\in \mathbb{N}\\
&\Rightarrow&\nonumber\\
\frac{4g_2 l_1+g_1}{2n_1+1}&=&\frac{2g_2-g_1}{k},\ k\in \mathbb{N}\label{eq:krestriction}\\
&\Rightarrow&\nonumber\\
C&=&\frac{2g_2-g_1}{2g_2 k}\label{eq:c1}\\
E^\infty&:=&\frac{q^2 A_\phi^2}{2m^*}\bigg(1-\bigg[\frac{2g_2-g_1}{2g_2 k}\bigg]^2\bigg)\label{eq:divene-1}
\end{eqnarray}
A similar argument applies for $2<g^*$, where we find an analogous expression, by using $n_2:=n_1-\Delta_n,\ \Delta n \in \mathbb{N}$ instead,
\begin{eqnarray}
C=\frac{|2g_2-g_1|}{2g_2 k}\label{eq:c2}
\end{eqnarray}
which eventually yields,
\begin{eqnarray}
E^\infty=\frac{q^2 A_\phi^2}{2m^*}\bigg(1-\bigg[\frac{|2-g^*|}{2 k}\bigg]^2\bigg),\ g^*\in \mathbb{Q}^*_+ \setminus \{2\}.\label{eq:divene-2}
\end{eqnarray}
So far we have not said anything about the allowed $k$-numbers except that they must be positive integers. However, there are additional constraints with respect to $g_1,g_2$ as well as the quantum numbers $n_1,l_1,n_2,l_2$ to be obeyed. From relation in Eq. (\ref{eq:krestriction}), we notice the following, depending on whether or not $g_1$ 
is an even $e\in\mathbb{E}:=\{2n,\ n \in \mathbb{N}\}$ or odd $o\in\mathbb{O}:=\{2n-1,\ n \in \mathbb{N}\}$ number:
\begin{eqnarray}
g_1=o,\ 2g_2= e \Rightarrow \frac{o_1}{o_2}=\frac{o_3}{k} \Rightarrow k \in\mathbb{O}\\
g_1=e,\ 2g_2= e \Rightarrow \frac{e_1}{o_2}=\frac{e_3}{k} \Rightarrow k \in\mathbb{O}\label{eq:evenodd}
\end{eqnarray}
Hence we have found that $k$ must be an odd number.
In a last step, we look at the special non-realtivistic case $g^* =2$ for which we find trivially,
\begin{eqnarray}
\frac{2l_1+1}{2n_1+1}&=&\frac{2l_1+1}{2n_2+1}\\ 
n_2&=&n_1,
\end{eqnarray}
which implies that Eq. (\ref{eq:divergence_relation}) holds almost everywhere expect for the lowest flat band,\cite{sidler2022class} due to the condition $l+g^* s/2>0$ for the angular quantum numbers. This inequality at the same time also imposes that
\begin{eqnarray}
C<1, 
\end{eqnarray}
which sets limits on the minimally allowed quantum numbers $k$, depending on $g_1$ and $g_2$ according to Eqs. (\ref{eq:c1}) \& (\ref{eq:c2}). The resulting expression for the squeezed Rydberg like series, imposed by Eq. (\ref{eq:divergence_relation}), is then given in Eq. (\ref{eq:rydbergdivergence}) of the main text.

\section{Derivation of an analytic solution for a time-dependent magnetic field in the slowly varying limit\label{app:analytic_solution}}

In order to solve the quasi-static eigenvalue problem of Eq. (\ref{eq:hamiltonianmc2}), we follow an analogous procedure to Ref. \citenum{sidler2022class} and choose a cylindrical coordinate system, which yields
\begin{eqnarray}
\hat{\tilde{H}}&=&\sum_{j=1}^N \bigg[-\frac{\hbar^2}{2m^*}\vec{\nabla}^2_j+ \frac{A_\phi q\hbar}{m^*}\bigg(\iu  \frac{\partial}{r_j\partial\phi_j} -\frac{g_s^*\sigma_{z,j}}{4 r_j}\bigg)
-\iu\frac{q\hbar}{8m^{*2} c^2 }\hbar\sigma_{z,j}\dot{A}_\phi\Big[\frac{1}{r_j}+2\frac{\partial}{\partial r_j}\Big]\bigg]+N\frac{q^2 A_\phi^2 }{2 m^*},\label{eq:clasham}\nonumber\\
\end{eqnarray}
 assuming a strictly two-dimensional electron gas.
Fortunately, the contribution of the diamagnetic term $ E_{A^2}:=\frac{q^2 A_\phi^2 }{2 m^*}$ remains constant for all $N$ electrons in radial coordinates, which reduces the complexity of our problem considerably. 
In a next step, we introduce
\begin{eqnarray}
 \alpha&:=&\frac{A_\phi q \hbar}{m^*}>0,  \\
 \epsilon&:=&\frac{q\hbar^2}{4m^{*2}  c^2}\dot{A}_\phi
\end{eqnarray}
which allows a more compact notation and the resulting eigenvalue problem for a single electron can be written as
\begin{eqnarray}
\bigg[-\frac{\hbar^2}{2m^*}\Big(\frac{\partial^2}{\partial r^2}+\frac{\partial}{r \partial r}+\frac{\partial^2}{r^2\partial \phi^2}\Big)+ \iu \alpha \frac{\partial}{r\partial\phi}+\frac{\alpha g^*\sigma_{z}}{4r}-\iu\epsilon s\Big[\frac{1}{r}+2\frac{\partial}{\partial r}\Big]\bigg]\tilde{\Psi}&=&\tilde{E}^-\tilde{\Psi}, \label{eq:sg}\nonumber\\
\end{eqnarray}
where the constant $E_{A^2}$-term is neglected for the moment. 
The angular and spin problem can trivially be solved by separation of variables as $\tilde{\Psi}(r,\phi,s)=\tilde{R}(r)\Phi(\phi)\chi(s)$, with spin function $\chi$ and $\Phi=e^{\iu l \phi}$ with $l\in Z$, $s=\pm\frac{1}{2}$, since $[\hat{\tilde{H}},\frac{\partial}{\partial\phi_j}]=0$. This leaves us with the radial problem
\begin{eqnarray}
\hat{\tilde{H}}_{l,s}R&:=&
\bigg[-\frac{\hbar^2}{2m^*}\big(\frac{\partial^2}{\partial r^2}+\frac{\partial}{r \partial r}-\frac{l^2}{r^2}\big)-\alpha \frac{l+g^* s/2}{r}-\iu\epsilon s\Big[\frac{1}{r}+2\frac{\partial}{\partial r}\Big]\bigg]\tilde{R}=\tilde{E}^- \tilde{R}.\label{eq:radial}
\end{eqnarray}

\textbf{Bound state eigenvalues:} 
%
%
To solve for the attractive eigenvalue problem,  we apply the method of Frobenius and match orders of a series expansion similar to Ref. \citenum{sidler2022class}. 
Therefore, we define 
\begin{eqnarray}
    \rho&:=&\sqrt{\frac{8 m^*|\tilde{E}^-|}{\hbar^2}}r,\label{eq:rhodef}\geq 0\\
    \lambda_{l,s}&:=&\alpha (l+g^* s/2) \sqrt{\frac{m^*}{2\hbar^2|\tilde{E}^-|}},\\
    \delta&:=&\epsilon\sqrt{\frac{m^* |\tilde{E}^-|}{2\hbar^2}},\ 0\leq\delta^2<1 
\end{eqnarray}
for which our radial problem assumes a convenient form, 
\begin{eqnarray}
\bigg[\frac{\partial^2}{\partial \rho^2}+\frac{\partial}{\rho \partial \rho}-\frac{l^2}{\rho^2}+  \frac{\lambda_{l,s}}{\rho}-\frac{1}{4}-\frac{\iu \delta s}{\rho}-2\iu\delta s\frac{\partial}{\partial\rho}\bigg]\tilde{R}(\rho)=0.\label{eq:radial_u}
\end{eqnarray}
To reach a simple closed form solution, we introduce the Ansatz 
\begin{eqnarray}
\tilde{R}(\rho)=e^{-\beta\rho} f(\rho)\label{eq:radialansatz}
\end{eqnarray}
with
\begin{eqnarray}
\beta:=-\iu \delta s+\frac{1}{2}\sqrt{1-4\delta^2 s^2}=-\iu \delta s+\frac{1}{2}\sqrt{1-\delta^2}.
\end{eqnarray}
This choice imposes an upper bound for $\delta^2$, i.e. $\delta^2<1$. This ensures that we have a normalizable solution for $\tilde{R}(\rho)$, given in Eq. (\ref{eq:radialansatz}), i.e. having an  exponentially decaying real part for $\rho\rightarrow\infty$.  Notice that large $\delta^2$-values would also imply a rapidly changing external vector potential, which certainly contradicts the initial assumption of a slow (adiabatic) time-evolution. Consequently, we find an ODE for $f(\rho)$ of the following form
\begin{eqnarray}
\bigg[\frac{\partial^2}{\partial \rho^2}-2\beta\frac{\partial}{\partial \rho}+\frac{\partial}{\rho \partial \rho}-\frac{l^2}{\rho^2}+  \Big(\lambda_{l,s}-\beta\Big)\frac{1}{\rho}-\frac{\iu \delta s}{\rho}-2\iu\delta s\frac{\partial}{\partial\rho}\bigg]f(\rho)=0.\label{eq:radial_f}
\end{eqnarray}
If we apply the series representation $f(\rho)=\sum_{j=0}^\infty c_j \rho^j$
and match the different orders in $\rho$, we find after an index shift $j\mapsto j+1$ with $c_{-1}=0$: 
\begin{eqnarray}
\sum_{j=-1}^\infty c_{j+1} j(j+1)\rho^{j-1}-2\beta c_j j\rho^{j-1}-2\iu\delta s c_j j\rho^{j-1}+c_{j+1} (j+1) \rho^{j-1} &&\\
- \beta c_j \rho^{j-1} -l^2 c_{j+1} \rho^{j-1}+\lambda_{l,s} c_j \rho^{j-1}-\iu \delta s c_j \rho^{j-1}&&=0\nonumber \label{eq:shiftedseries}
\end{eqnarray}
This gives  rise to the indicial equation:
\begin{eqnarray}
c_{j+1}[(j+1)^2-l^2]=c_j[\sqrt{1-\delta^2}j+\frac{\sqrt{1-\delta^2}}{2}-\lambda_{l,s}].
\end{eqnarray}
It implies the "series switches on" for $c_{j+1}$ when $(j+1)^2=l^2$, i.e. $j+1=l$, and it can terminate only if $\sqrt{1-\delta^2(|\tilde{E}^-|)}j+\sqrt{1-\delta^2(|\tilde{E}^-|)}/2-\lambda_{l,s}(\tilde{E}^-) =0$. Otherwise one would converge to a non-normalizable solution since $c_{j+1}\rightarrow \frac{c_j}{j}$ for large $j$ and $f\rightarrow \sum_{j=0}^\infty\frac{\rho^j}{j !}$. This also implies that $\lambda>0$, i.e. we find a bound state solution only for $l+g^*s/2>0$ similar to the strictly static results.\cite{sidler2022class} Now, introducing quantum number $n:=j$ leads to a quadratic equation in $\tilde{E}^-$:
\begin{eqnarray}
 \Big(n+\frac{1}{2}\Big)^2=\frac{\lambda^2}{1-\delta^2}= \frac{\alpha^2(l+g^* s/2)^2 m^*}{2\hbar^2}\frac{1}{|\tilde{E}^-|}\Bigg[\frac{1}{1-\frac{\epsilon^2 m^*}{2\hbar^2}|\tilde{E}^-|}\Bigg],\ n\geq l. \label{eq:nqnr}
\end{eqnarray}
 with a simple closed form solution for the energy eigenvalues 
\begin{eqnarray}
\tilde{E}^-_{n,l,s}=\pm\frac{\hbar^2}{m^* \epsilon^2}\Bigg[1\pm\sqrt{1-\frac{\alpha^2\epsilon^2m^{*2}
}{\hbar^4}\bigg[\frac{2l+g^* s}{2n+1}\bigg]^2}\Bigg],\ n\geq l,\ l+g^* s/2>0.\label{eq:Ebound}
\end{eqnarray}
We will subsequently see that the quantum numbers $n,l,s$ uniquely define an eigenfunction of the PDE given in Eq. (\ref{eq:sg}) with corresponding unique eigenvalue $\tilde{E}$. To determine the correct sign, we notice that for $\epsilon\rightarrow 0$ our PDE given in  Eq. (\ref{eq:sg}) reduces to the problem of a static vector potential ($\dot {\vec{A}}_\phi=0$) with known solutions.\cite{sidler2022class} The known eigenvalues are indeed recovered exactly by choosing the negative signs twice, i.e., from
\begin{eqnarray}
    E_{n,l,s}^-&\overset{!}=&\lim_{\epsilon\rightarrow 0} \tilde{E}^-_{n,l,s}= \lim_{\epsilon\rightarrow 0}\Bigg\{-\frac{\hbar^2}{m^* \epsilon^2}\Bigg[1-\sqrt{1-\frac{\alpha^2\epsilon^2m^{*2}
}{\hbar^4}\bigg[\frac{2l+g^* s}{2n+1}\bigg]^2}\Bigg]\Bigg\}\\
&=&-\frac{ q^2 A_{\phi}^2
}{2m^*}\bigg[\frac{2l+g^* s}{2n+1}\bigg]^2,
\end{eqnarray}
whereas a positive sign in front of the square-root would lead to a divergent limit instead.
Finally, we obtain Eqs. (\ref{eq:etilde})-(\ref{eq:etilde_e}), as introduced in the main part of the manuscript.

\textbf{Eigenfunctions:}
After having identified the bound state energy eigenvalues for $n\geq l, l+g^* s/2>0$, we can next find the corresponding eigenfunctions by expressing $f(\rho) = \rho^{l} L(\rho)$.\cite{schlicht} This turns Eq.~\eqref{eq:radial_f} into
\begin{eqnarray}
\rho \frac{d ^2 L}{d\rho^2}+(2l+1-\sqrt{1-\delta^2}\rho)\frac{dL}{d\rho}+\bigg[-\frac{\sqrt{1-\delta^2}}{2}-l\sqrt{1-\delta^2}+\lambda\bigg] L&=&0,\label{eq:lagidrho}
\end{eqnarray}
which can be further simplified by substituting $\tilde{x}:=\sqrt{1-\delta^2}\rho$ and using Eq. (\ref{eq:nqnr}) to,
\begin{eqnarray}
\tilde{x} \frac{d ^2 L}{d\tilde{x}^2}+(2l+1-\tilde{x})\frac{dL}{d\tilde{x}}+(n-l) L&=&0,\ n,l\in \mathbb{N}_0.\label{eq:lagid}
\end{eqnarray}
This PDE can be solved by the associated Laguerre polynomials $L_{n-l}^{2l}$ of degree $n-l$ and parameter $2l$.~\cite{arfken1999mathematical} The associated Laguerre polynomials are given by Rodrigues' formula,\cite{arfken1999mathematical}
\begin{eqnarray}
L_w^\nu(\tilde{x})=\frac{\tilde{x}^{-\nu} e^{\tilde{x}}}{w!}\frac{d^w}{d\tilde{x}^w}(e^{-\tilde{x}}\tilde{x}^{w+\nu}).
\end{eqnarray}
Therefore, the radial solution of our problem given in Eq. (\ref{eq:radial}) becomes
\begin{eqnarray}
\tilde{R}_{n,l,s}(\tilde{x})=e^{-\iu \frac{\delta s}{\sqrt{1-\delta^2}} \tilde{x}}e^{-\frac{\tilde{x}}{2}}\bigg(\frac{\tilde{x}}{\sqrt{1-\delta^2}}\bigg)^l L_{n-l}^{2l}(\tilde{x}).\label{eq:eigu}
\end{eqnarray}
Consequently, the orthonormal eigenfunctions of the full problem are found as given in Eqs. (\ref{eq:tildeeigf}) - (\ref{eq:tilde_norm}) in the main section. 
The orthogonality of the eigenfunctions can be shown identically to the argument given in Ref. \citenum{sidler2022class}, where surprisingly the same degeneracy argument remains valid because of Eq. (\ref{eq:etilde_e}).



\end{document}